\documentclass{sig-alternate} 
\usepackage{mathptmx} 

\newcommand{\ignore}[1]{}
\usepackage{fancyhdr}
\usepackage[normalem]{ulem}
\usepackage[hyphens]{url}
\usepackage{microtype}
\usepackage{color}
\usepackage{epsfig}
\usepackage{epstopdf}
\usepackage{graphicx}
\usepackage{multirow}
\usepackage{subcaption}
\usepackage{wrapfig}
\usepackage{amsmath}
\usepackage{dsfont}
\usepackage{comment}
\usepackage{authblk}

\fancypagestyle{firstpage}{
  \fancyhf{}

  \pagenumbering{arabic}
}  

\title{TRINITY: Coordinated Performance, Energy and Temperature Management in 3D Processor-Memory Stacks} 
\author[1]{Karthik Rao}
\author[2]{William Song}
\author[3]{Yorai Wardi}
\author[4]{Sudhakar Yalamanchili} 
\affil[1]{Georgia Institute of Technology, Atlanta, USA \\ 
			raokart@gatech.edu}
\affil[2]{Yonsei University, Seoul, South Korea \\ 
	wjhsong@yonsei.ac.kr}
\affil[1]{Georgia Institute of Technology, Atlanta, USA \\ 
	ywardi@ece.gatech.edu}
\affil[1]{Georgia Institute of Technology, Atlanta, USA \\ 
	sudha@gatech.edu}

\begin{document}
\maketitle
\thispagestyle{firstpage}
\pagestyle{plain}

\begin{abstract}
The consistent demand for better performance has lead to innovations at hardware and microarchitectural levels. 3D stacking of memory and logic dies delivers an order of magnitude improvement in available memory bandwidth. The price paid however is, tight thermal constraints. 

In this paper, we study the complex multiphysics interactions between performance, energy and temperature. Using a cache coherent multicore processor cycle level simulator coupled with power and thermal estimation tools, we investigate the interactions between (a) thermal behaviors (b) compute and memory microarchitecture and (c) application workloads. The key insights from this exploration reveal the need to manage performance, energy and temperature in a coordinated fashion. Furthermore, we identify the concept of ``effective heat capacity" i.e. the heat generated beyond which no further gains in performance is observed with increases in voltage-frequency of the compute logic. Subsequently, a real-time, numerical optimization based, application agnostic controller (TRINITY) is developed which intelligently manages the three parameters of interest. We observe up to $30\%$ improvement in Energy Delay$^2$ Product and up to $8$ Kelvin lower core temperatures as compared to fixed frequencies. Compared to the \texttt{ondemand} Linux CPU DVFS governor, for similar energy efficiency, TRINITY keeps the cores cooler by $6$ Kelvin which increases the lifetime reliability by up to 59\%.
\end{abstract}

\section{Introduction}
\label{sec:Intro}
The performance of data intensive computing systems that process terabytes of data is increasingly limited by data movement and corresponding energy overheads. 3D packaging technologies enabled by advances such as Through-Silicon-Via (TSV) technology \cite{motoyoshi2009through}, has led to stacking of silicon dies thereby enabling the integration of memory and logic in a small footprint with significant reductions in data movement latency and energy. Further, the package provides an order of magnitude increase in memory bandwidth. For example, commercial standards like DDR3-1333 \cite{ddr3jedec}, DDR4-2667 \cite{ddr4jedec}, HBM2 \cite{hbmjedec} and HMC2 \cite{hmcconsortium} realize 10.66 GB/s, 21.34 GB/s, 256 GB/s and 320 GB/s, respectively. To effectively exploit the high bandwidth provided by 3D die stacked DRAM, multiple efforts have explored moving compute logic inside the package as part of the die stack revisiting the early efforts at architecting Processing-In-Memory (PIM) designs \cite{banerjee20013,liu2005bridging,black2006stacking,loh20083d,woo2010optimized,borkar20113d,emma20143d}. The compute logic layer in the 3D stack can range from simple atomic operations to multiple Out-of-Order (OoO) cores to general purpose low power GPUs. However, stacking memory and logic dies in this manner exacerbates thermal effects which, if left unchecked will preclude any performance gains from co-locating compute and memory. In particular, the exponential relationship between temperature and leakage current diminishes the performance that can be achieved for the heat capacity of the package thereby limiting the ability to exploit the order of magnitude increase in available memory bandwidth \cite{eckert2014thermal,milojevic2012thermal}. 
\begin{figure}[!t]
    \centering
    \includegraphics[width=0.45\textwidth]{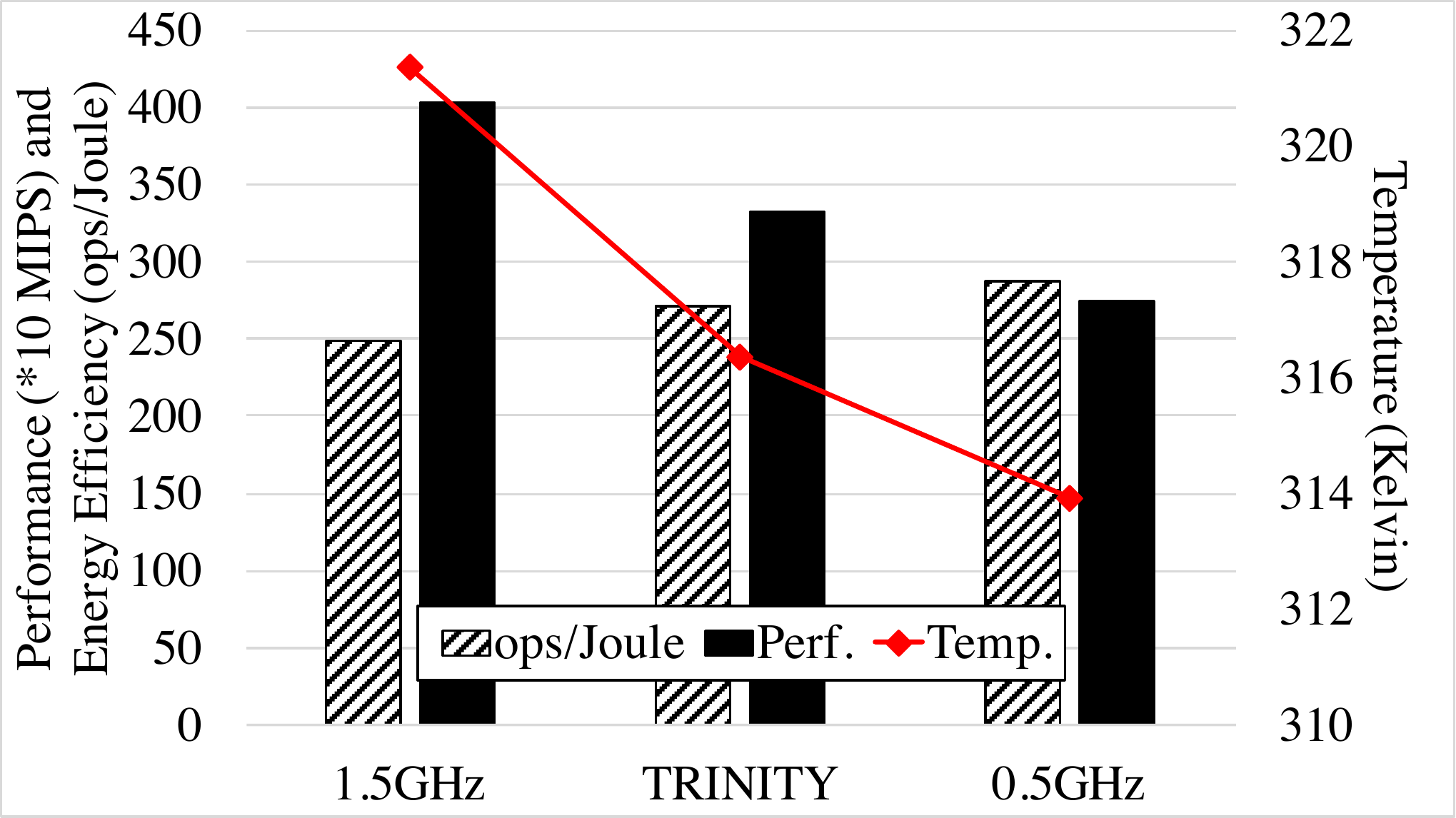}
    \vspace{-2mm}
    \caption{\textbf{TRINITY balancing performance, energy and temperature by effectively utilizing the heat capacity}.}
    \label{fig:Figure1}
    \vspace{-6mm}
\end{figure}

There is a rich body of work on managing thermal effects in processors. Software-based efforts \cite{ahn2014dynamic,khurshid2013data,lo2016thermal,zhao2013temperature,tran2013heterogeneous} typically seek to redistribute heat to avoid peak temperature violations. Hardware based efforts employ dynamic voltage frequency scaling (DVFS) to manage the thermal fields  \cite{meng2012optimizing,chen2015thermal,kang2011maximizing}. Detailed thermal modeling using software packages such as HotSpot \cite{zhang2015hotspot} and 3D-ICE \cite{sridhar20103d} enables the study of microarchitectural effects on temperature. Although bulk of the work has been pursued for 2D packages, the understanding is still relevant to 3D packages. For example, researchers have explored the thermal coupling between cores on the same layer and between cores on different vertically stacked layers \cite{zhu2008three}. In general these approaches have dealt with temperature as a constraint. We argue that temperature is a resource that has to be managed, like memory or compute cycles. This approach is rooted in a different view of the relationship between performance and heat capacity. 




The heat capacity of the package is established based on the thermal design power (TDP) which is set independent of the application characteristics. However, some applications such as sparse matrix computations have components that are memory bound rather than compute bound. Temperature-based approaches to improve the performance of such applications by boosting voltage-frequency in an attempt to utilize the thermal headroom \cite{rotem2012power}, will simply waste power with little or no performance gain and significant reductions in energy efficiency. An example is shown in Figure \ref{fig:Figure1} where energy efficiency(temperature) of a memory bound benchmark decreases(increases) with increasing clock frequency. On the other hand, compute intensive applications such as dense matrix algebra may extract performance benefits from DVFS schemes but can exceed the temperature bounds. Furthermore, thermal coupling between adjacent cores can increase leakage current (and therefore static power) and accelerate temperature rise leading to premature throttling \cite{paul2013cooperative} and therefore performance and energy efficiency loss. 

Our goal is to ensure that for the amount of heat generated by the compute logic for an application, the maximum performance (throughput) is delivered. Doing so must implicitly improve energy efficiency. A key insight is that applications, and some application phases, simply cannot utilize the package thermal headroom even when operating at the highest voltage-frequency state. We attempt to capture this observation by noting that for a specific application or phase there is an \textit{effective heat capacity} (EHC) - this is the heat generated beyond which further gains in performance are in-feasible with further increases in voltage-frequency of the compute logic. For example, an application may be operating in a memory bound phase and increases in compute logic frequency has little effect on performance but may consume the thermal headroom. Accordingly, we note that the EHC is application-specific and time-varying. Consequently, our goal is to maximize the performance that can be extracted from the time-varying EHC. The solution must be online, adaptive, and robust to modeling errors. The EHC corresponds to a value of temperature which we will refer to as the {\em effective maximum temperature (EMT)}. Practical implementations will seek to operate at the EMT and minimize thermal coupling induced leakage power. 

Therefore, we propose TRINITY, a DVFS controller that implements an on-line optimization technique that continuously balances performance, energy efficiency and thermal behaviors to fully utilize the EHC (See Fig. \ref{fig:Figure1}). Each voltage island implements an independently operated TRINITY controller that is (i) based on numerical optimization, (ii) is computationally inexpensive to implement, (iii) is self-tuning, (iv) distributed (per-core), and (iv) application agnostic. The behavior of spatially adjacent controllers is implicitly coupled via temperature. Thus a network of interacting controllers locally seeks to maximize throughput from the locally available EHC and their coordinated actions indirectly makes the most efficient use the package heat capacity. Our vehicle for exploration and demonstration is a cache-coherent multicore processor integrated as the bottom die in a 3D DRAM stack as shown in the Figure \ref{fig:SystemPhysicalLayout}. Cores operate on distinct voltage islands and are capable of operating in independent power states each with a TRINITY controller. The controller is designed considering practical implementation challenges such as (i) measurement and actuation delays (ii) computation delays and (iii) hardware limitations such as having a discrete set of voltage-frequency states. In cycle-level simulations of a 16 core architecture, for $10\%$ increase in Energy Delay Product, TRINITY keeps the temperature lower by $6$ Kelvin while achieving similar energy efficiency as compared to \texttt{ondemand} Linux CPU governor. An added benefit of the reduced temperature is the increase in lifetime reliability of the 3D stack by up to 59\%.\\

\noindent This paper seeks to make the following contributions:
\begin{enumerate}
    \item The introduction of the concept of effective heat capacity as a thermal resource to be managed. (Section \ref{sec:BG})
	\item Development of TRINITY, an on-line DVFS controller (i) based on numerical optimization, (ii) computationally inexpensive to implement, (iii) self-tuning, (iv) distributed (per-core), and (v) application agnostic for cooperatively balancing the performance, energy efficiency, and thermal behaviors of applications. (Section \ref{sec:Optimization})
	\item A comprehensive simulation-based characterization of intra- and inter-die thermal coupling effects demonstrating the need to maximally utilize the effective heat capacity. (Section \ref{sec:Characterization})
\end{enumerate}


\section{Motivation}
\label{sec:BG}
\vspace{-4mm}
\begin{figure}[!htb]
    \centering
    \includegraphics[scale=0.3]{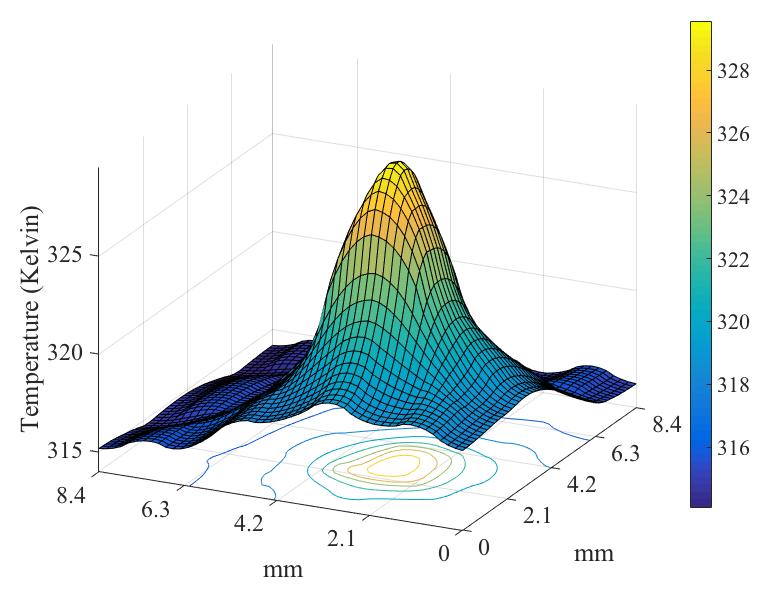}
    \vspace{-2mm}
    \caption{\textbf{Heat map of the core layer showing reduction in thermal headroom for neighboring cores}.}
    \label{fig:HeatMap}
    \vspace{-4mm}
\end{figure}
Consider a 3D architecture as described Section \ref{sec:Framework} and illustrated in Figures \ref{fig:SystemFunctional} and \ref{fig:SystemPhysicalLayout}, where 16 cores are integrated at the bottom of a 3D DRAM stack. When only one of the cores is executing an application thread while the rest are idle, the resulting thermal gradient from the `hot' core to the neighboring `cool' cores is shown in Figure \ref{fig:HeatMap}. We note that the program thread executing on a core can increase the temperature of neighboring cores by $> 7$ Kelvin. Not shown here, is another observation that on migrating this thread from a location next to the package boundary to a location in the center of the core die decreases the temperature by up to $10$ Kelvin (these are computed as steady state temperatures). {\em Ideally, we would like the thermal gradients to be zero, performance to be maximum, and the temperature to be the local EMT at every core.} 

\begin{figure}[!htb]
    \centering
    \includegraphics[width=0.45\textwidth,height=2.7in]{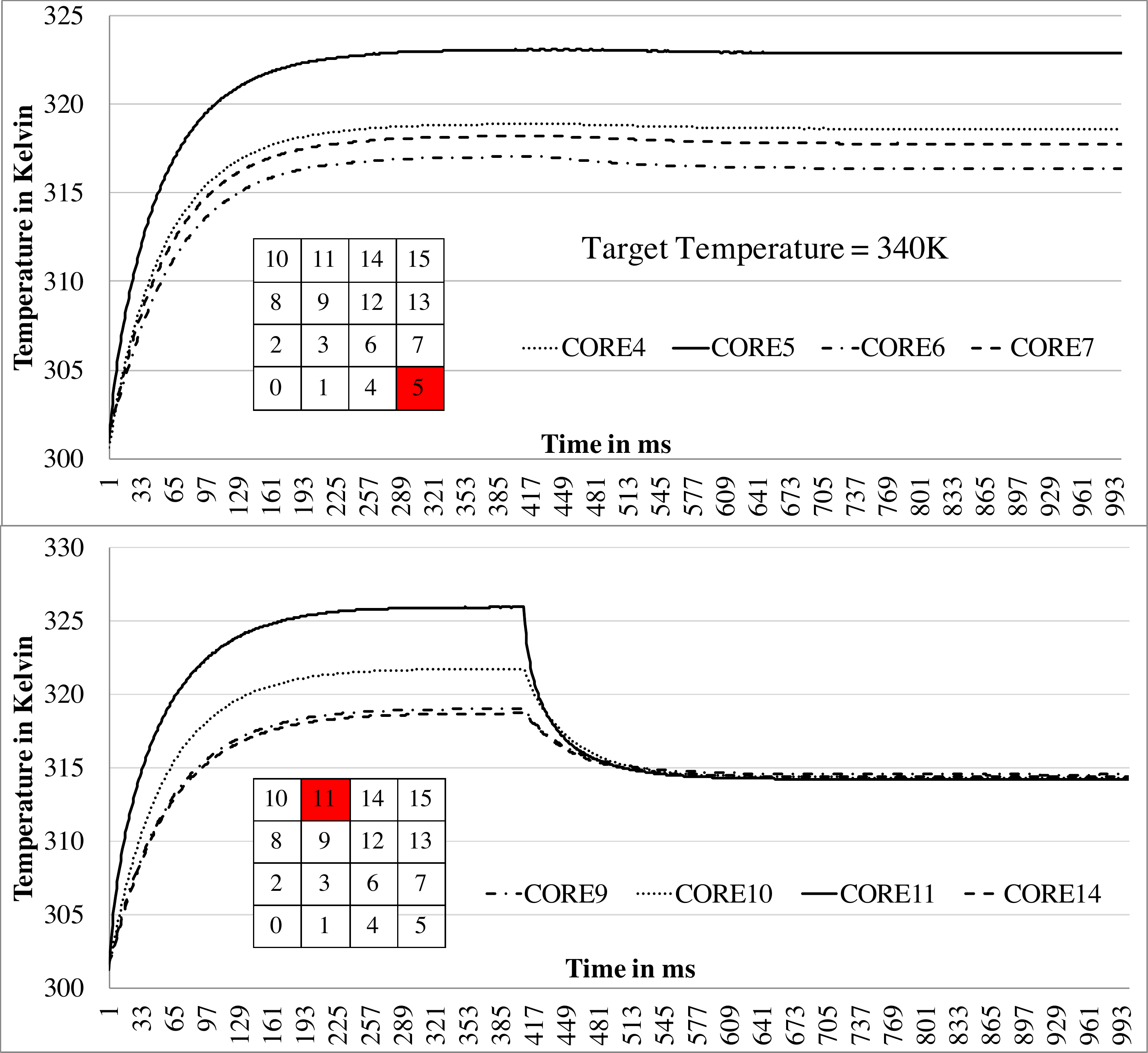}
    \vspace{-2mm}
    \caption{\textbf{Temperature Regulation Inefficiency: Except for $\text{Core}_5$ and $\text{Core}_{11}$, rest of the cores are idle. At $400$ms mark $\text{Core}_{11}$ becomes idle as well}.}
    \label{fig:Regulation1}
    \vspace{-2mm}
\end{figure}

Achieving this goal via temperature regulation techniques are of limited utility. For example, consider the use of a temperature regulator~\cite{rao2015temperature} at each core. The objective of the regulator is to maintain a fixed temperature. We ran a graph benchmark and set the target temperature to $340$ Kelvin for each core. In Figure~\ref{fig:Regulation1} we observe that none of the cores can reach the target temperature. For cores which are idle i.e. threads are waiting to be woken up, the controller tries to raise the temperature of the core by increasing the corresponding voltage-frequency but ends up wasting energy through increases in leakage power due the rise in temperature and no improvement in performance of the core. Temperature regulation in this form is therefore inefficient in 3D stacks because, (1) target tracking temperature (which is the EMT) has to be known \textit{apriori}, (2) target temperature will be different for different cores and will vary at run-time, and (3) temperature dynamics is a rather slow process (100s of milli-seconds) in comparison to application characteristics that can vary rapidly (micro-seconds). {\em Therefore control techniques must be on-line, adaptive, and application agnostic.}

The preceding example with temperature regulation illustrates an important point - for certain applications and during certain application phases, package heat capacity cannot be utilized completely. This points to the existence of an {\em effective heat capacity} which roughly corresponds to the temperature of the cores beyond which there is little or no increase in performance. We refer to this temperature as the {\em effective maximum temperature}. It also represents an energy efficient (ops/joule) operating point. Note that the heat capacity of the entire package is established independent of the specific workload and that effective heat capacity of an application can be time varying. A thread currently in a memory intensive phase with EMT of $X$ Kelvin, may transition into a compute intensive phase where its EMT is $Y$ Kelvin ($Y > X$). Without profiling an application extensively, tracking the EMT is a challenge. To further emphasize the effect of EHC, we present data from two benchmark applications \texttt{blackscholes} (PARSEC \cite{bienia2008parsec}) and \texttt{tc} (GraphBig \cite{nai2015graphbig}) in Table \ref{tab:EffectiveThermalCapacity}. The average temperature of the cores in Kelvin and average performance (Million-Instructions-Per-Second (MIPS)) for the two benchmarks is listed for three different fixed frequencies.

\vspace{-2mm}
\begin{table}[!htb]
    \centering
     \caption{\textbf{Table demonstrating variable application heat capacities and room for improving balance between performance, temperature and energy}.}
    \label{tab:EffectiveThermalCapacity}
    \vspace{-2mm}
    \resizebox{0.47\textwidth}{!}{
    \begin{tabular}{|p{1.7cm}||c|c|c|c|}
        \hline
                                              & \textbf{Bench.} & \textbf{0.5GHz} & \textbf{1.0GHz} & \textbf{1.5GHz} \\ \hline \hline
        \multirow{2}{*}{Temp. (K)} & \texttt{blacks.} & $318.18$ & $329.68$ & $340.93$  \\ \cline{2-5}
                                              & \texttt{tc} & $313.91$ & $318.93$ & $323.85$ \\ \hline
        \multirow{2}{*}{Perf. (MIPS)}   & \texttt{blacks.} & $12378.3$ & $23710.7$ & $33065.6$  \\ \cline{2-5}
                                              & \texttt{tc} & $2741.3$ & $3633.9$ & $4026.7$ \\ \hline 
        \multirow{2}{*}{ED2P}                 & \texttt{blacks.}    & $0.67$   & $0.33$   & $0.26$   \\ \cline{2-5}
                                              & \texttt{tc}             & $0.76$   & $0.58$   & $0.58$  \\ \hline 
    \end{tabular}
    }
    \vspace{-2mm}
\end{table}
Performance and temperature characteristics of both applications vary widely. In order to demonstrate that there is room to improve performance, energy and temperature in these systems, we also compute the Energy $\text{Delay}^2$ Product (ED2P) at the three fixed frequencies. For compute intensive applications like \texttt{blackscholes}, best ED2P is achieved at the highest frequency. But, for memory intensive benchmarks like \texttt{tc}, there is no appreciable improvement in ED2P beyond $1.0$GHz, {\em The goal of TRINITY is to dynamically track these behaviors with distributed on-line control. }

We observe that it is important to make a distinction between peak temperature and effective heat capacity. The former is a constraint that all thermal management schemes seek to observe. Heat capacity reflects the net amount of heat that can be generated. Observing only the former will not maximize performance for the corresponding amount of heat. Our goal is to extract as much performance as possible from the heat generated by the application. Thread scheduling techniques that seek to redistribute heat can be re-purposed towards this end. {\em In this sense, effective heat capacity is a resource which the proposed on-line distributed controller network is designed to exploit efficiently}.

In light of the observations made in the previous paragraphs, we first present a microbenchmark characterization of the 3D stacked architecture in Section \ref{sec:Characterization}. We list the key insights which lead us to development of the optimization problem and its solution is described in Section \ref{sec:DTM}. We then proceed to evaluate the proposed controller over a set of benchmark applications in Section \ref{sec:Results}. Finally, we list relevant works in Section \ref{sec:RelatedWork} and conclude the paper in Section \ref{sec:Conclude}.
\section{Characterization}
\label{sec:Characterization}
In this section we seek to find answers to the following questions:\\
\noindent (1) What is the thermal impact of a hot core on neighboring cool cores? What are the performance implications for both the hot core and the cool cores? \\
\noindent (2) What is the thermal and performance behavior of a program thread executing at different physical locations on the core layer? \\
\noindent (3) How does memory addressing patterns in the L2 Cache layer affect the temperature of the core layer and vice versa?\\ 

We therefore proceed by characterizing temperature and performance of the 3D stack (See Section \ref{sec:Framework}, Fig. \ref{fig:SystemFunctional} and \ref{fig:SystemPhysicalLayout}) under a variety of microbenchmark workloads. Temperature is measured in Kelvin and performance in MIPS. The temperature numbers reported are steady state values. The microbenchmarks are designed such that they (i) exhibit variable ops/byte ratio, (ii) access specific memory locations, and (iii) execute on specific physical cores. 
\subsection{Nomenclature}
\label{sec:Nomenclature}

\begin{figure*}[!htb]
	\centering
    \includegraphics[scale=0.5]{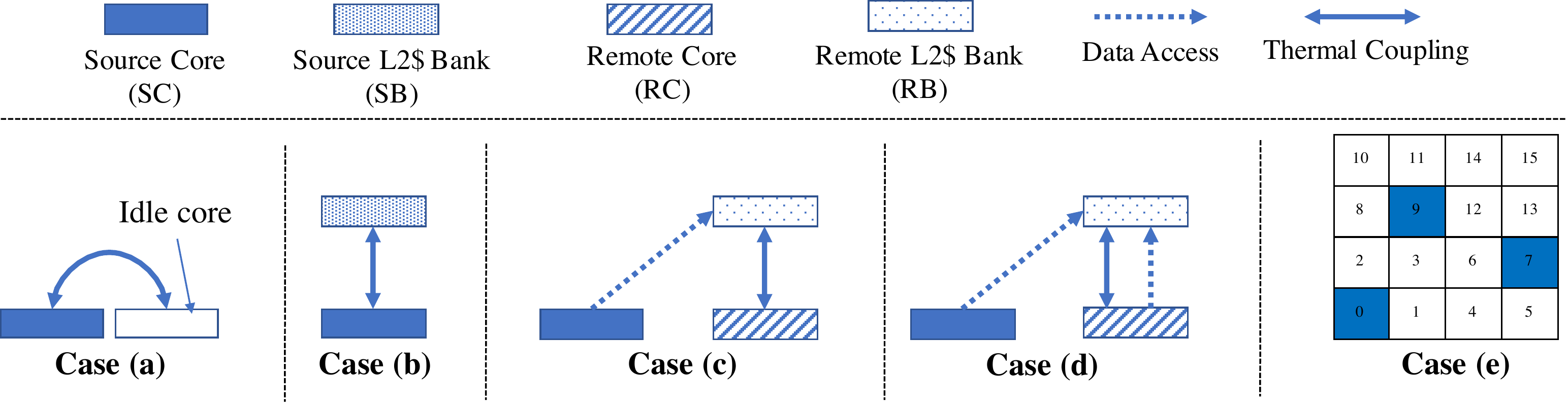}
    \vspace{-2mm}
	\caption{\textbf{Microbenchmark characterization nomenclature}.}
	\label{fig:Nomenclature}
	\vspace{-2mm}
\end{figure*}

To better represent the characterization results, we first describe a naming convention in Figure \ref{fig:Nomenclature} which is used throughout the characterization section of this paper. All the microbenchmarks are single threaded programs. Most of the results that follow have a single thread running on a single fixed core (source core) accessing data from a single fixed L2 Cache bank (source/remote bank). We make a distinction when two cores are running independent microbenchmark applications as and when required. While a microbenchmark is running on a single core, the rest of the cores are powered up ($V_{dd}$ and $CLK$ are supplied) but idle. The 1-hop and 2-hop neighbors of the source core are termed SC+1 and SC+2, respectively. Similarly, for the L2 Cache banks we have SB+1 and SB+2. In what follows, a ``memory intensive benchmark" continuously performs \texttt{load} operations on sequential memory locations whereas a ``compute intensive benchmark" repeats the following two steps: (i) \texttt{load} a block of data from memory (ii) perform integer and floating point operations for a fixed number of iterations. 

Among the many cases of thermal coupling, we discuss 5 types in detail as shown in Figure \ref{fig:Nomenclature}: 
\begin{enumerate}
    \item [\textbf{(a)}] Thermal coupling between adjacent cores.
    \item [\textbf{(b)}] Thermal coupling between a core and an L2 Cache bank directly on top.
    \item [\textbf{(c)}] Thermal coupling between an L2 Cache bank and an idle core below it.
    \item [\textbf{(d)}] Same as \textbf{(c)} but with a non-idle core.
    \item [\textbf{(e)}] Thermal coupling variation when the computation is moved from the package boundary to the center of the die.
\end{enumerate}

\subsection{Thermal Coupling Analysis}
\label{sec:TCouplingAnalysis}
\begin{figure*}
	\begin{subfigure}[!htb]{0.3\textwidth}
		\includegraphics[width=2.3in,height=1.5in]{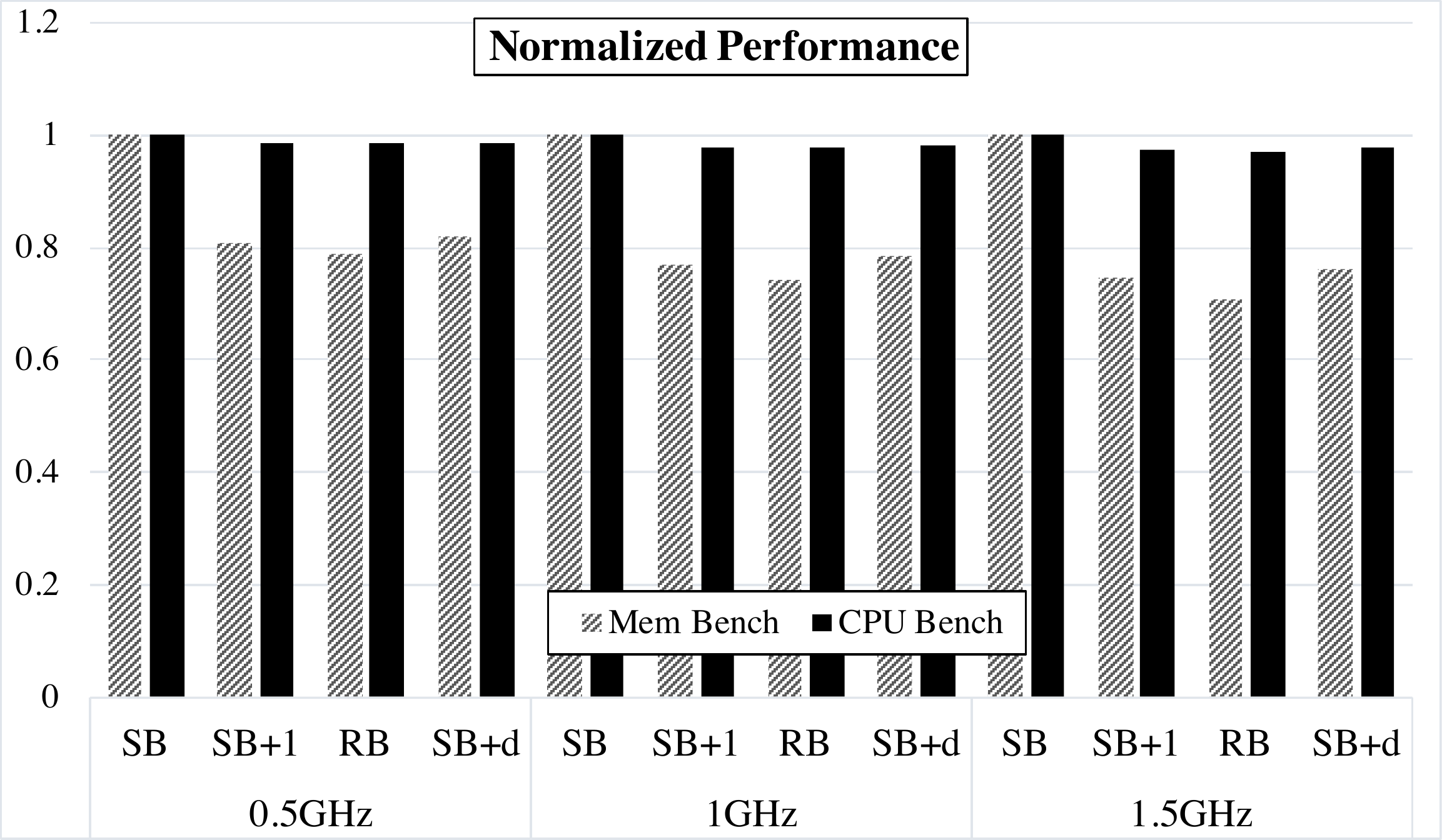}
		\caption{\textbf{Performance on y-axis is normalized w.r.t SB}}
		\label{fig:Core0Perf}
	\end{subfigure}
	~
	~
	~
	\begin{subfigure}[!htb]{0.3\textwidth}
		\includegraphics[width=2.3in,height=1.5in]{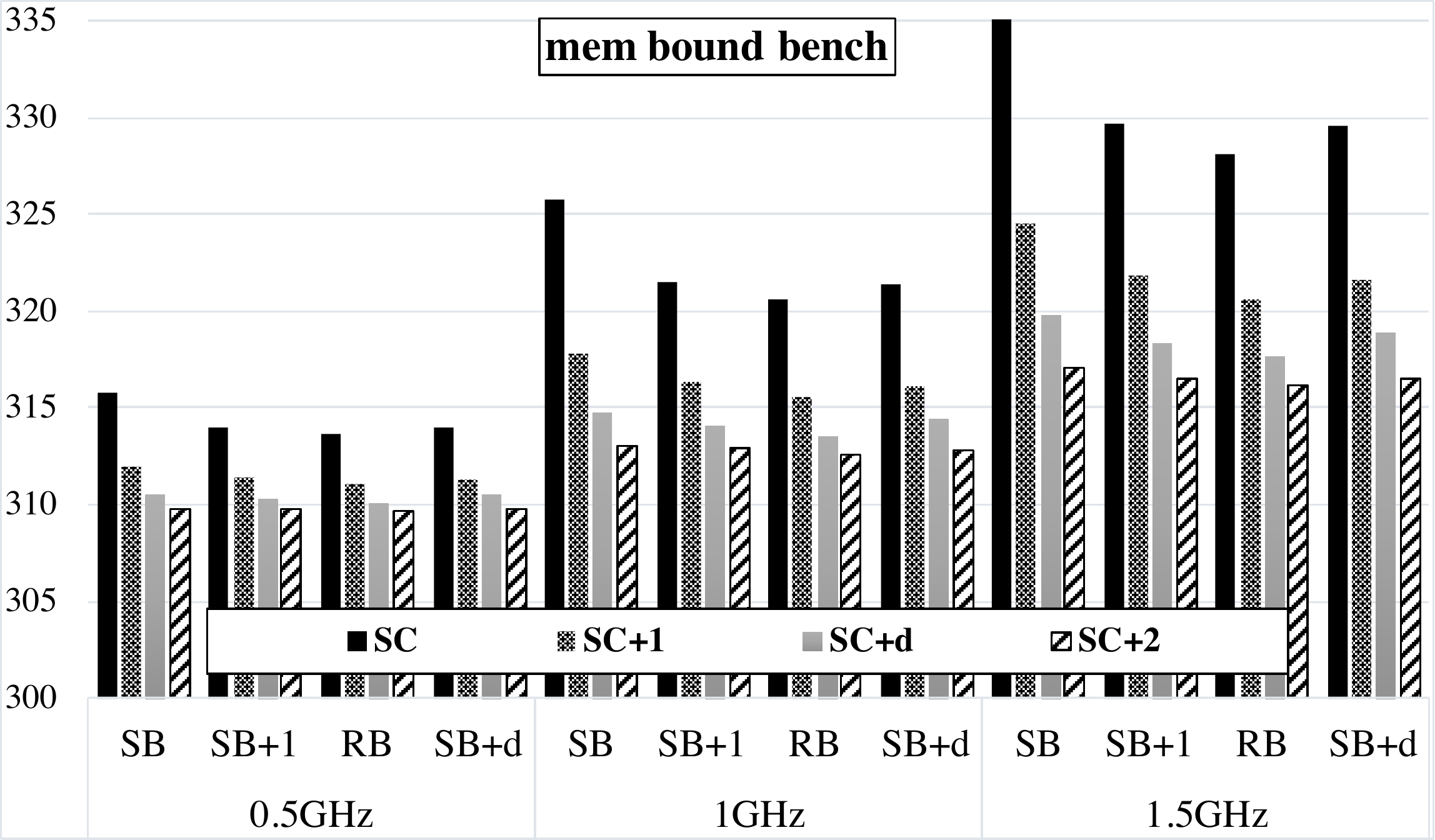}
		\caption{\textbf{Temperature of source core and its neighbors in Kelvin on y-axis}.}	
		\label{fig:Core0MemTemp}
	\end{subfigure}
	~
	~
	~
	\begin{subfigure}[!htb]{0.3\textwidth}
		\includegraphics[width=2.3in,height=1.5in]{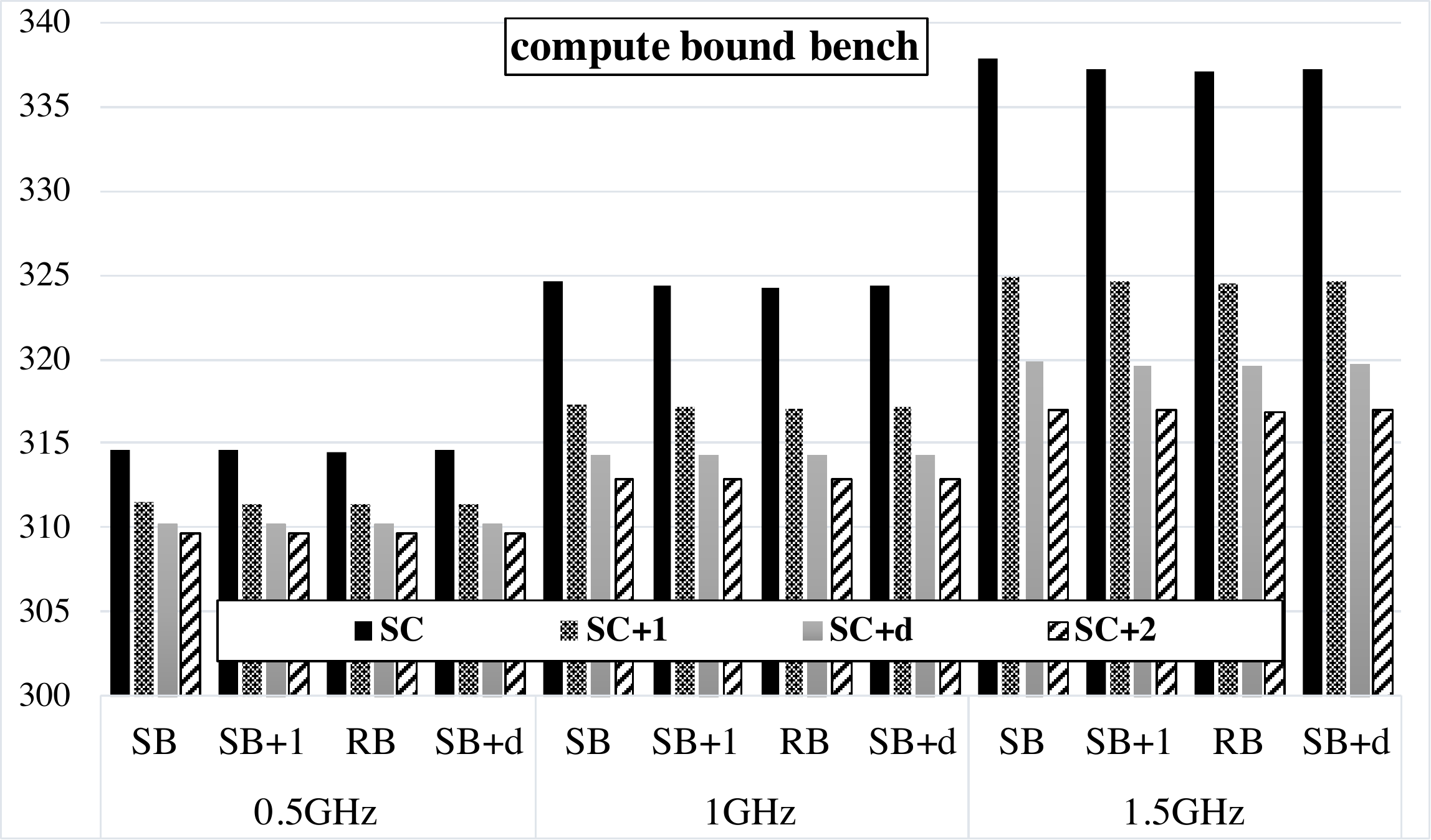}
		\caption{\textbf{Temperature of source core and its neighbors in Kelvin on y-axis}.}
		\label{fig:Core0CpuTemp}
	\end{subfigure}
	\vspace{-2mm}
	\caption{\textbf{Performance and temperature variation when running mem bound and compute bound benchmarks on a source core accessing source and remote cache banks at different core frequencies. SC is Core$_0$}.}
	\label{fig:Core0PerfTemp}
	\vspace{-4mm}
\end{figure*}
\begin{figure*}
	\begin{subfigure}[!htb]{0.3\textwidth}
		\includegraphics[width=2.3in,height=1.5in]{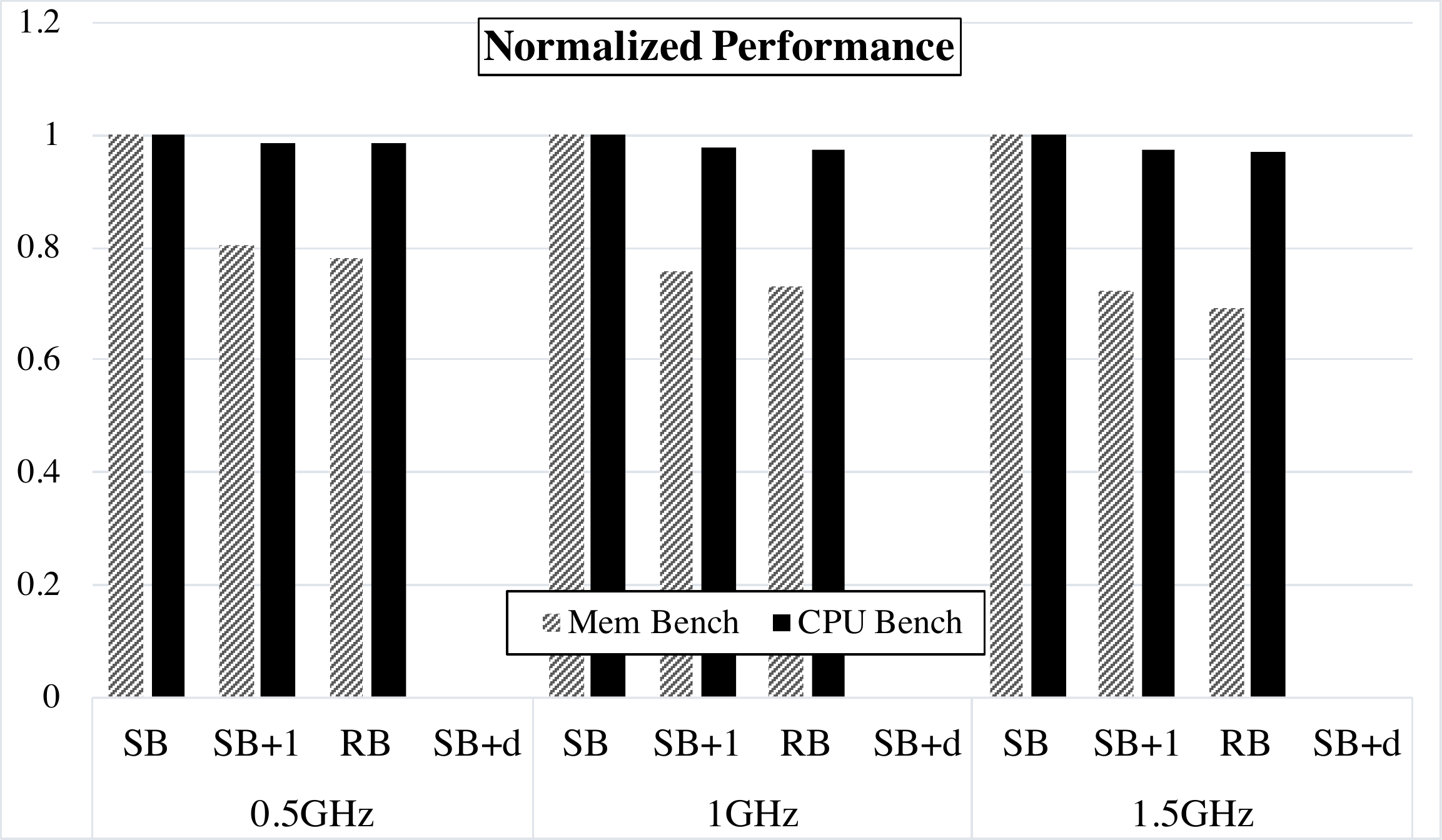}
		\caption{\textbf{Performance on y-axis is normalized w.r.t SB}}
		\label{fig:Core2Perf}
	\end{subfigure}
	~
	~
	~
	\begin{subfigure}[!htb]{0.3\textwidth}
		\includegraphics[width=2.3in,height=1.5in]{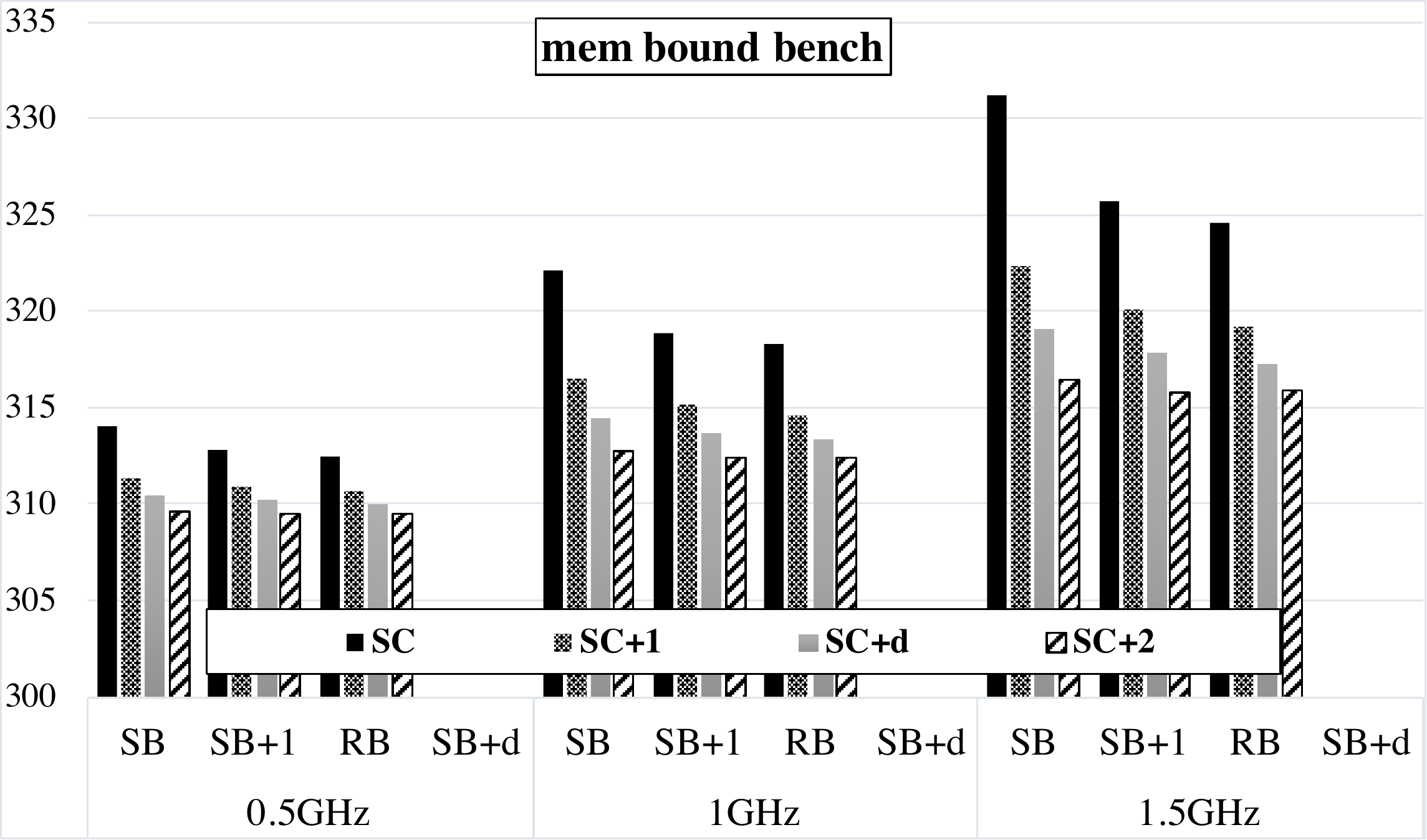}
		\caption{\textbf{Temperature of source core and its neighbors in Kelvin on y-axis}.}	
		\label{fig:Core2MemTemp}
	\end{subfigure}
	~
	~
	~
	\begin{subfigure}[!htb]{0.3\textwidth}
		\includegraphics[width=2.3in,height=1.5in]{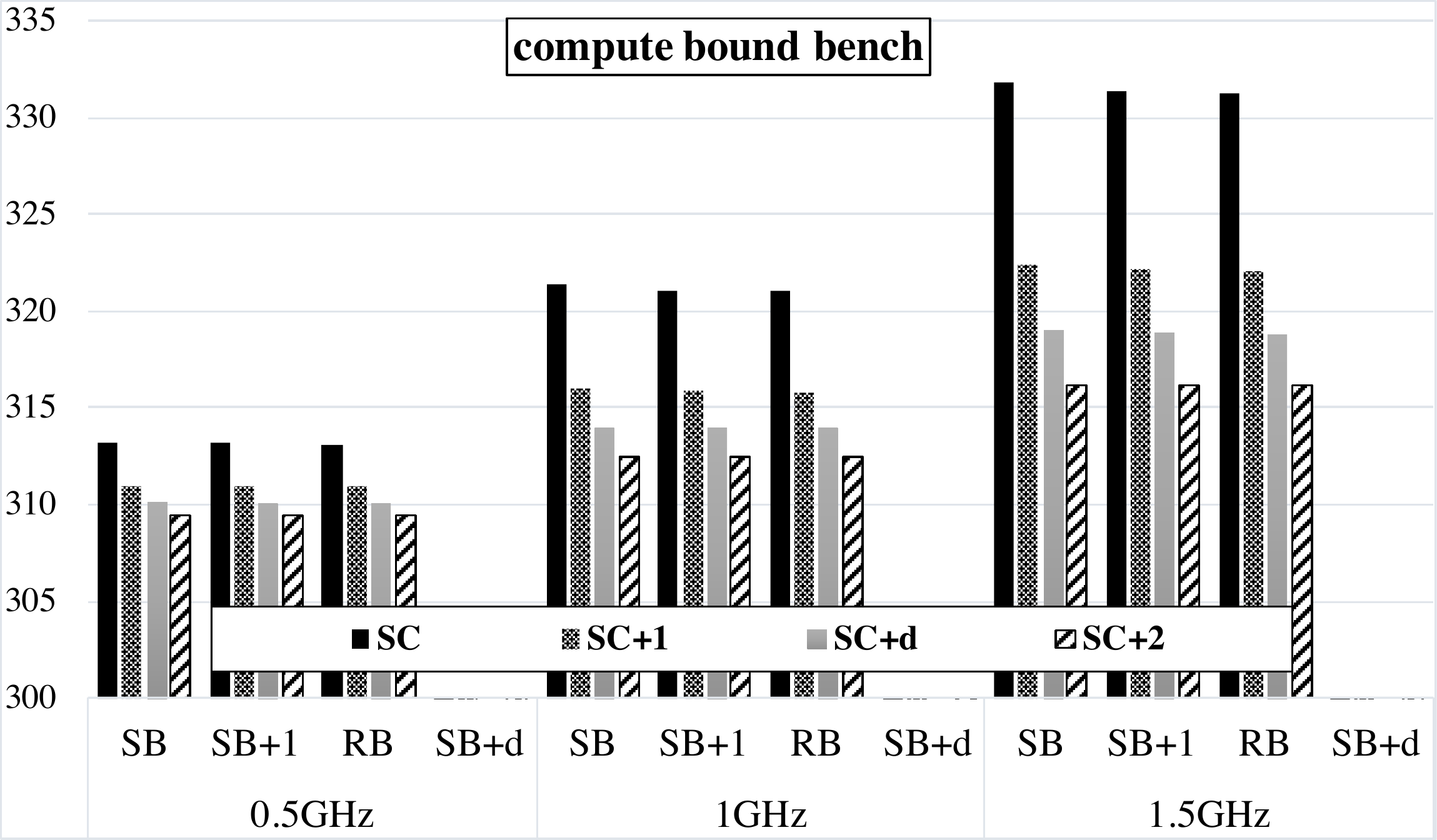}
		\caption{\textbf{Temperature of source core and its neighbors in Kelvin on y-axis}.}
		\label{fig:Core2CpuTemp}
	\end{subfigure}
	\vspace{-2mm}
	\caption{\textbf{Performance and temperature variation when running mem bound and compute bound benchmarks on a source core accessing source and remote cache banks at different core frequencies. SC is Core$_2$}.}
	\label{fig:Core2PerfTemp}
	\vspace{-4mm}
\end{figure*}
\begin{figure*}
	\begin{subfigure}[!htb]{0.3\textwidth}
		\includegraphics[width=2.3in,height=1.5in]{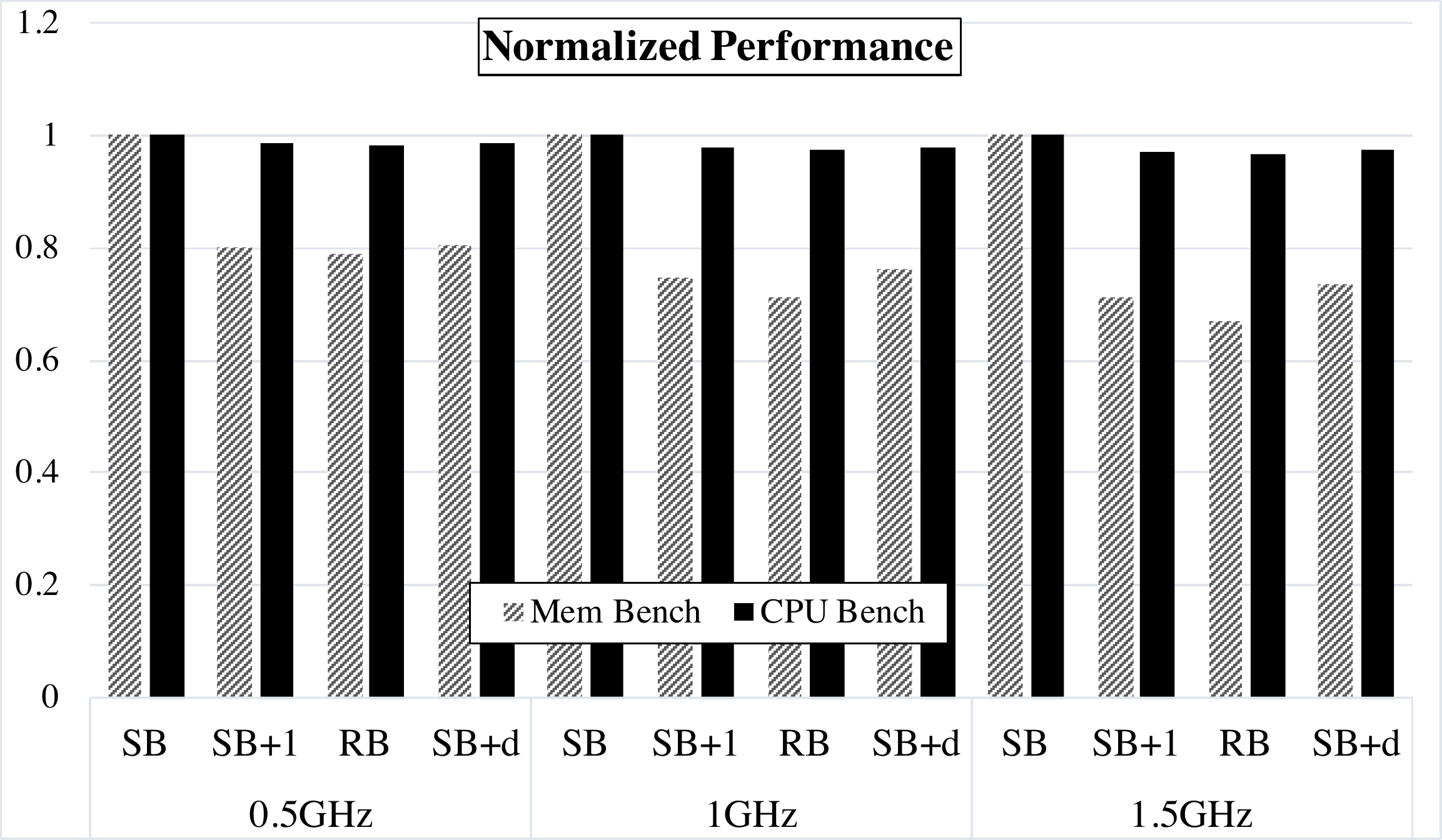}
		\caption{\textbf{Performance on y-axis is normalized w.r.t SB}}
		\label{fig:Core3Perf}
	\end{subfigure}
	~
	~
	~
	\begin{subfigure}[!htb]{0.3\textwidth}
		\includegraphics[width=2.3in,height=1.5in]{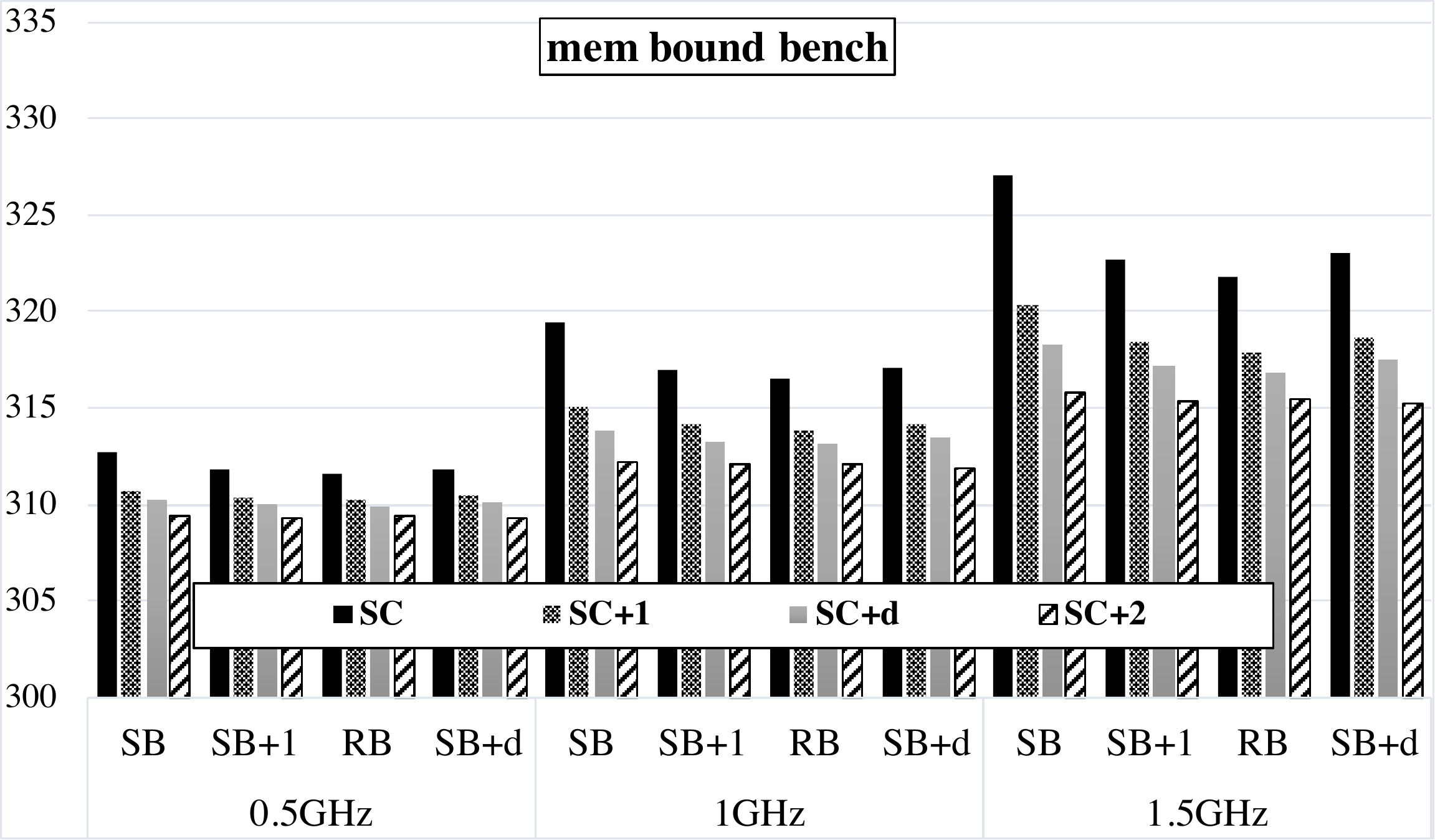}
		\caption{\textbf{Temperature of source core and its neighbors in Kelvin on y-axis}.}	
		\label{fig:Core3MemTemp}
	\end{subfigure}
	~
	~
	~
	\begin{subfigure}[!htb]{0.3\textwidth}
		\includegraphics[width=2.3in,height=1.5in]{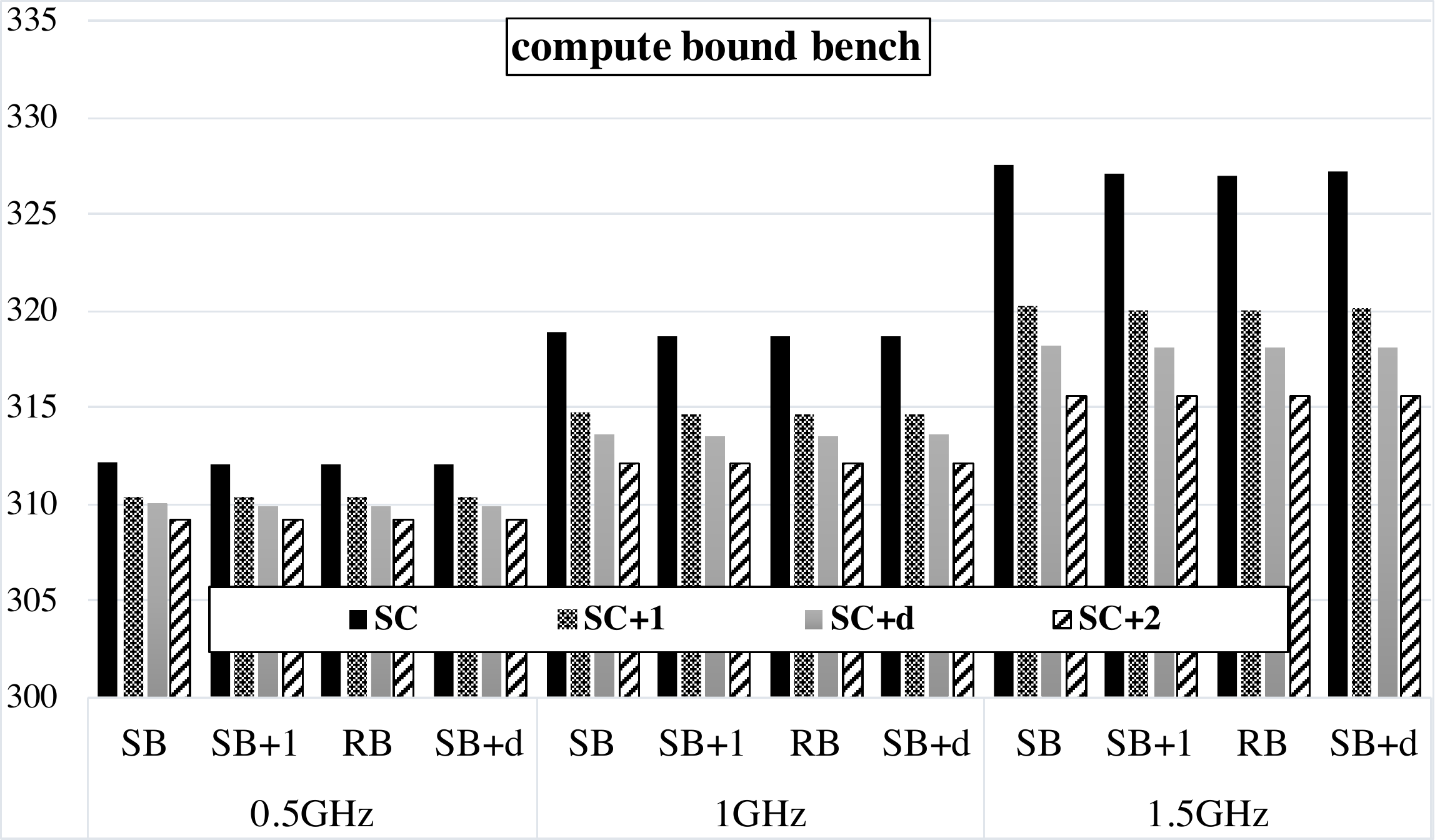}
		\caption{\textbf{Temperature of source core and its neighbors in Kelvin on y-axis}.}
		\label{fig:Core3CpuTemp}
	\end{subfigure}
	\vspace{-2mm}
	\caption{\textbf{Performance and temperature variation when running mem bound and compute bound benchmarks on a source core accessing source and remote cache banks at different core frequencies. SC in Core$_3$}.}
	\label{fig:Core3PerfTemp}
	\vspace{-4mm}
\end{figure*}
\noindent \textbf{Case (a):} In Figure \ref{fig:Core0MemTemp}, the temperature of the source core SC, average temperature of its 1-hop neighbors and 2-hop neighbors, SC+1 and SC+2, respectively is plotted for a memory bound microbenchmark at three different clock frequencies with the SC accessing data from SB and RB. Figure \ref{fig:Core0CpuTemp} is similar to Figure \ref{fig:Core0MemTemp} except that the microbenchmark is compute bound. Out of the 16 possible locations for the SC, accounting for symmetry, three locations viz. Core$_0$, Core$_2$ and Core$_3$ are chosen. The temperature and performance trends for Core$_0$, Core$_2$ and Core$_3$ are plotted in Figures \ref{fig:Core0PerfTemp}, \ref{fig:Core2PerfTemp} and \ref{fig:Core3PerfTemp}, respectively. We note that regardless of whether the benchmark running on SC is memory intensive or compute intensive, although SC+1 is idle, due to thermal coupling, steady state temperature of SC+1 can go as high as $325$ Kelvin. Thermal coupling effects are negligible beyond a 2-hop neighborhood concurring with prior work \cite{bartolini2013thermal} (albeit \cite{bartolini2013thermal} is for a 2D architecture). The extent of thermal coupling in a 3D architecture however, is more pronounced within the 1-hop neighborhood due to heat shielding from upper layers. The key observation we make is:\\
\textit{\underline{Observation 1}: A `hot' core reduces the EHC of neighboring `cool' cores by up to $7$ Kelvin.}

A second more subtle observation to obtained by analyzing the steady state temperature and performance of the SC when accessing SB and RB (See Fig. \ref{fig:Core0Perf} and \ref{fig:Core0MemTemp}). By addressing a RB, the SC temperature can be reduced by up to $8$ Kelvin. This however, comes at the price of $30\%$ reduction in performance. Therefore, \\
\textit{\underline{Observation 2}: Memory address re-mapping has the potential to trade-off performance for reduction in temperature.}\\



\begin{figure*}
	\centering
	\begin{subfigure}[!htb]{0.45\textwidth}
		\includegraphics[width=3.15in,height=1.5in]{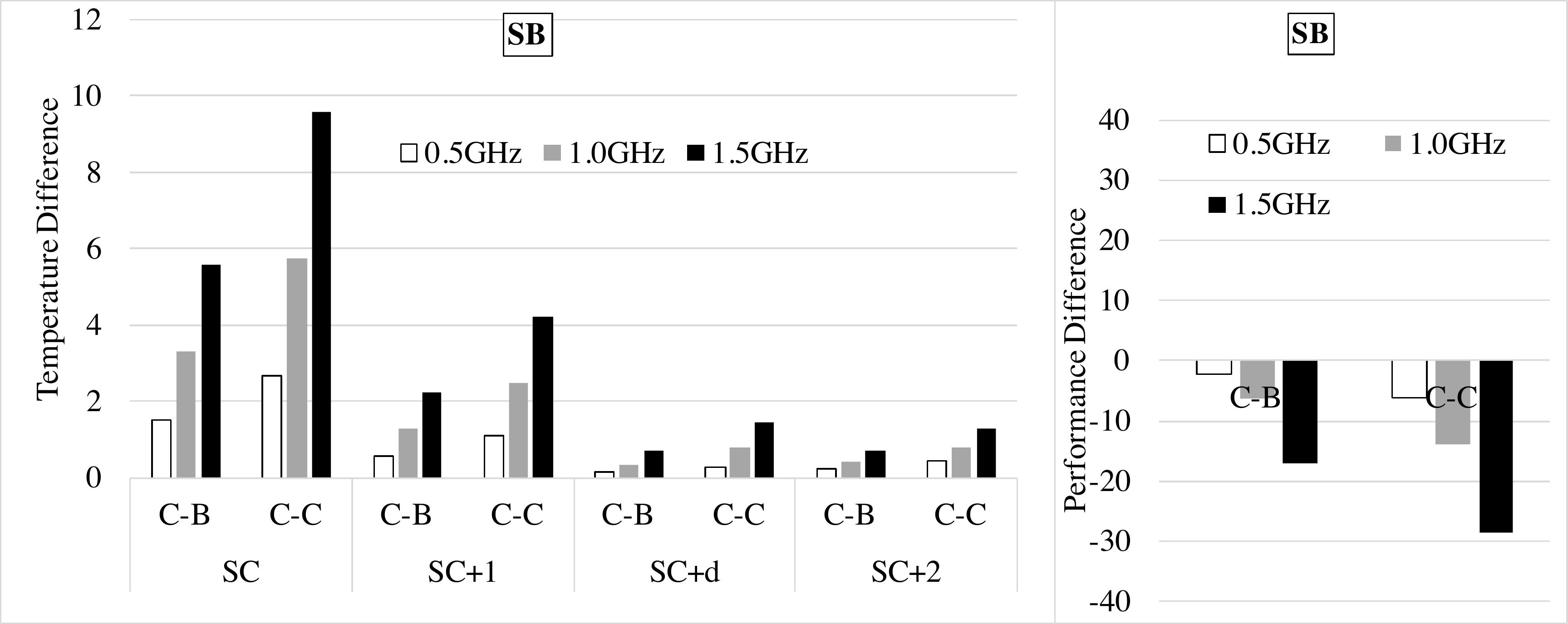}
		\caption{\textbf{Source core accessing source bank. Performance and spatial temperature comparison of source core at Corner vs. Center vs. Boundary}.}
    	\label{fig:CBCCPlotSB}
	\end{subfigure}
	~
	\begin{subfigure}[!htb]{0.45\textwidth}
		\includegraphics[width=3.15in,height=1.5in]{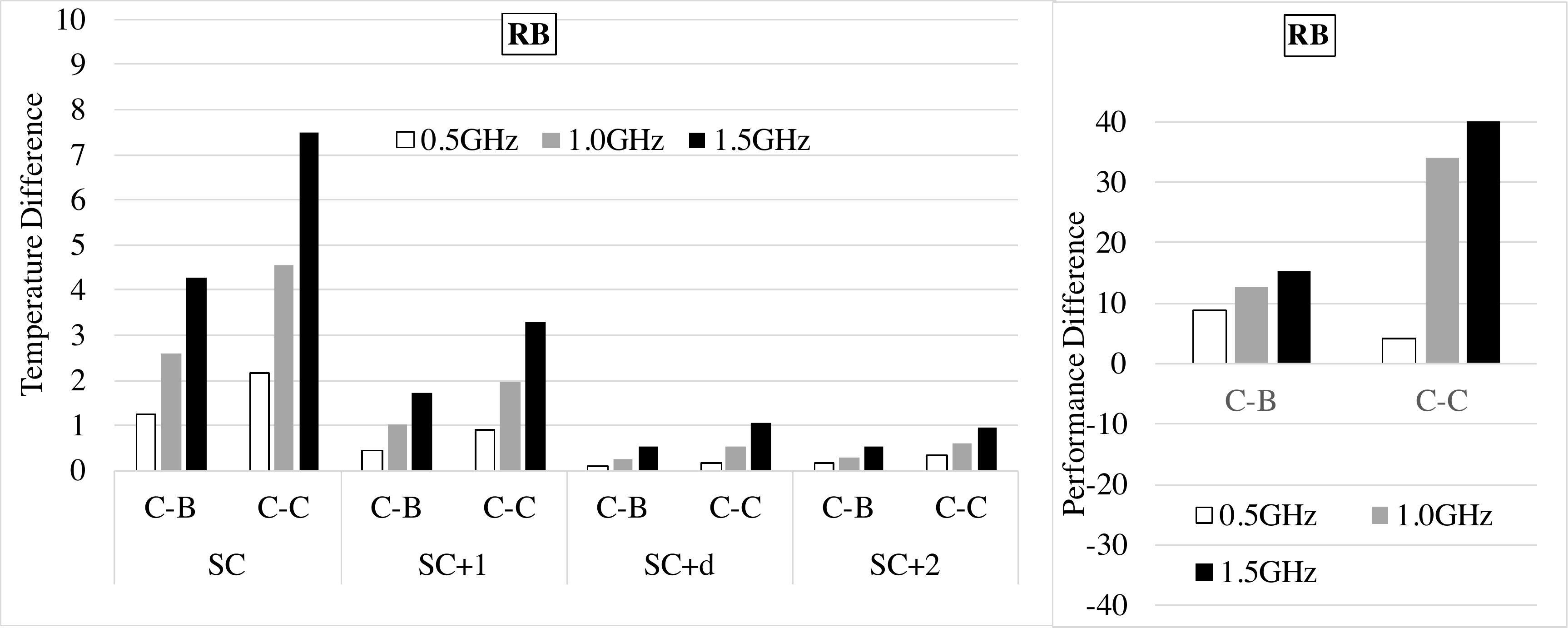}
		\caption{\textbf{Source core accessing remote bank. Performance and spatial temperature comparison of source core at Corner vs. Center vs. Boundary}.}
    	\label{fig:CBCCPlotRB}
	\end{subfigure}
	\vspace{-2mm}
	\begin{subfigure}[!htb]{0.45\textwidth}
		\includegraphics[width=3.15in,height=1.5in]{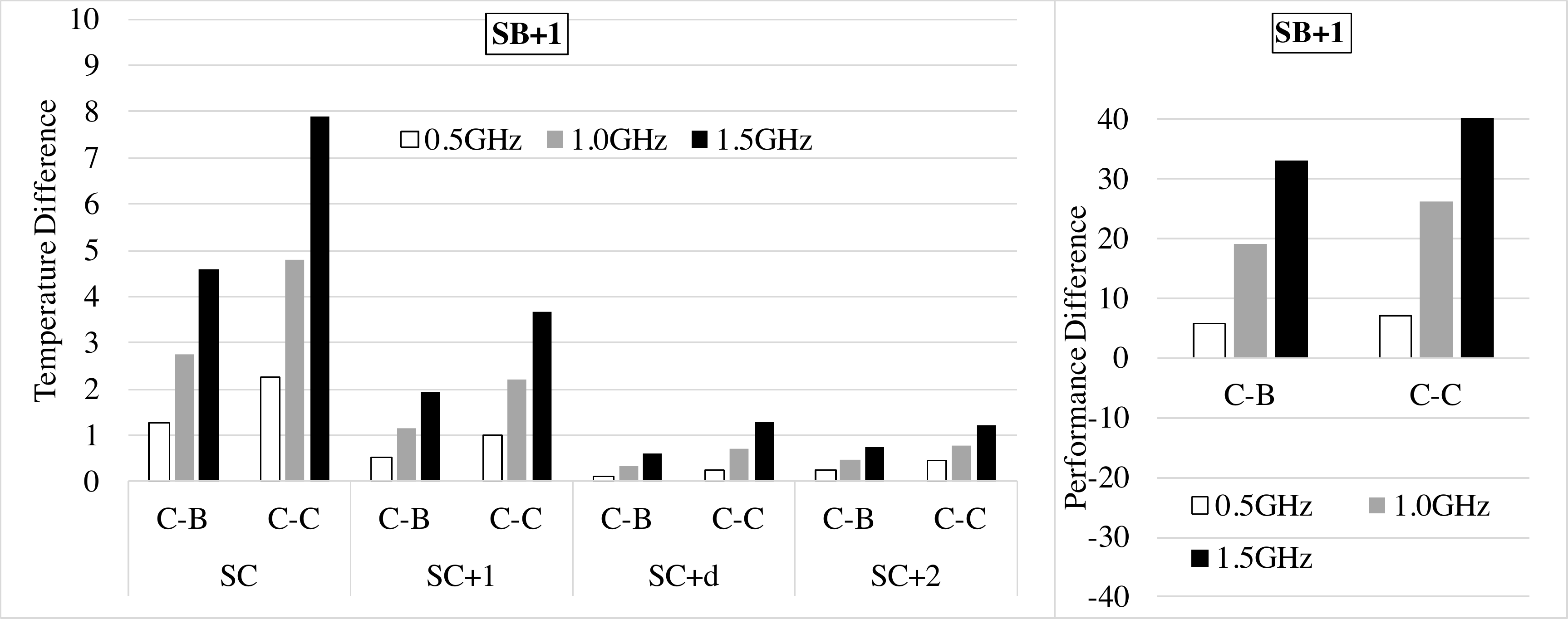}
		\caption{\textbf{Source core accessing 1-hop neighbor of source bank. Performance and spatial temperature comparison of source core at Corner vs. Center vs. Boundary}.}
    	\label{fig:CBCCPlotSB1}
	\end{subfigure}
	\caption{\textbf{Influence of package boundaries on thermal coupling and performance}.}
	\label{fig:CBCCAllPlots}
	\vspace{-4mm}
\end{figure*}

\noindent \textbf{Case (e):} Carrying forward from \textbf{Case (a)}, we repeat the same set of experiments but choose to run the microbenchmark on a SC that is physically located at three specific locations: (1) Corner (Core$_0$) (2) Boundary (Core$_2$ and (3) Center (Core$_3$). Using `Corner' as the reference, we compute the differences in temperature and performance for the other two locations. Specifically, we annotate the differences as follows: Corner - Boundary (C-B) and Corner - Center (C-C). The trend of the data obtained is plotted in Figure \ref{fig:CBCCAllPlots}. The difference between figures \ref{fig:CBCCPlotSB}, \ref{fig:CBCCPlotRB} and \ref{fig:CBCCPlotSB1} is only with the memory location addressed, SB, RB and SB+1, respectively.

Temperature difference in Kelvin and performance difference in MIPS are plotted on the y-axis. In general, moving the application thread from the corner to the boundary or center reduces the temperature of the SC between $1 - 10$ Kelvin with negligible loss in performance. The greatest difference is seen for the C-C case. Not only does the SC experience reduction in temperature, its neighbors SC+1 too benefit by up to $4$ Kelvin due to the relocation. Note however, that this phenomenon does not nullify \textbf{Case (a)}. Only the magnitude of thermal coupling is mitigated to a small extent. 

\textit{\underline{Observation 3}: Package boundaries become increasingly important in 3D stacked environments. OS level thread scheduling in cooperation with DVFS schemes can lead to better utilization of the EHC. }

\begin{figure}[!htb]
    \centering
    \begin{subfigure}[!htb]{0.45\textwidth}
		\includegraphics[width=3.5in,clip,trim={1.5cm 0 0 0}]{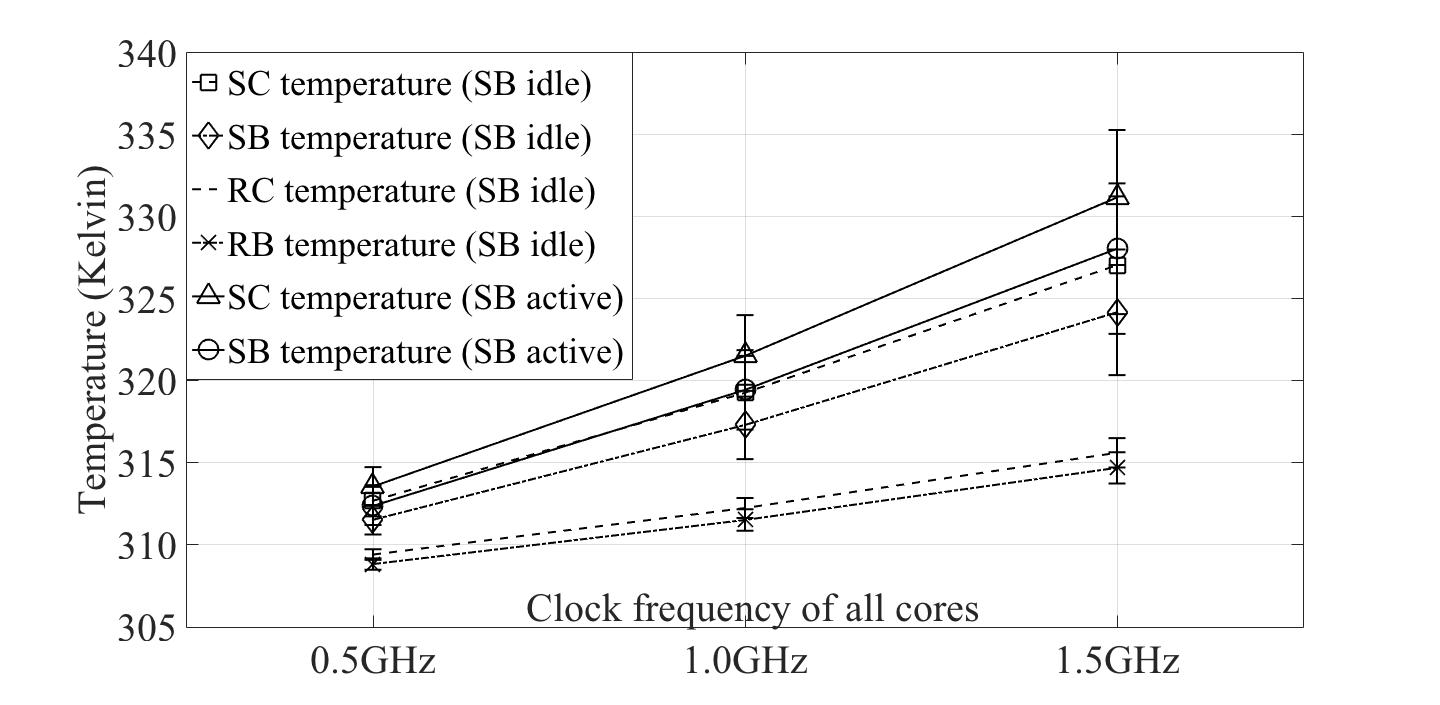}
		\caption{\textbf{Temperature variation of (i) source core and cache (ii) remote core and cache}.}
		\label{fig:InterLayerTemperature}
	\end{subfigure}
	\vspace{-2mm}
	\begin{subfigure}[!htb]{0.45\textwidth}
		\includegraphics[width=3.5in,clip,trim={1.5cm 0 0 0}]{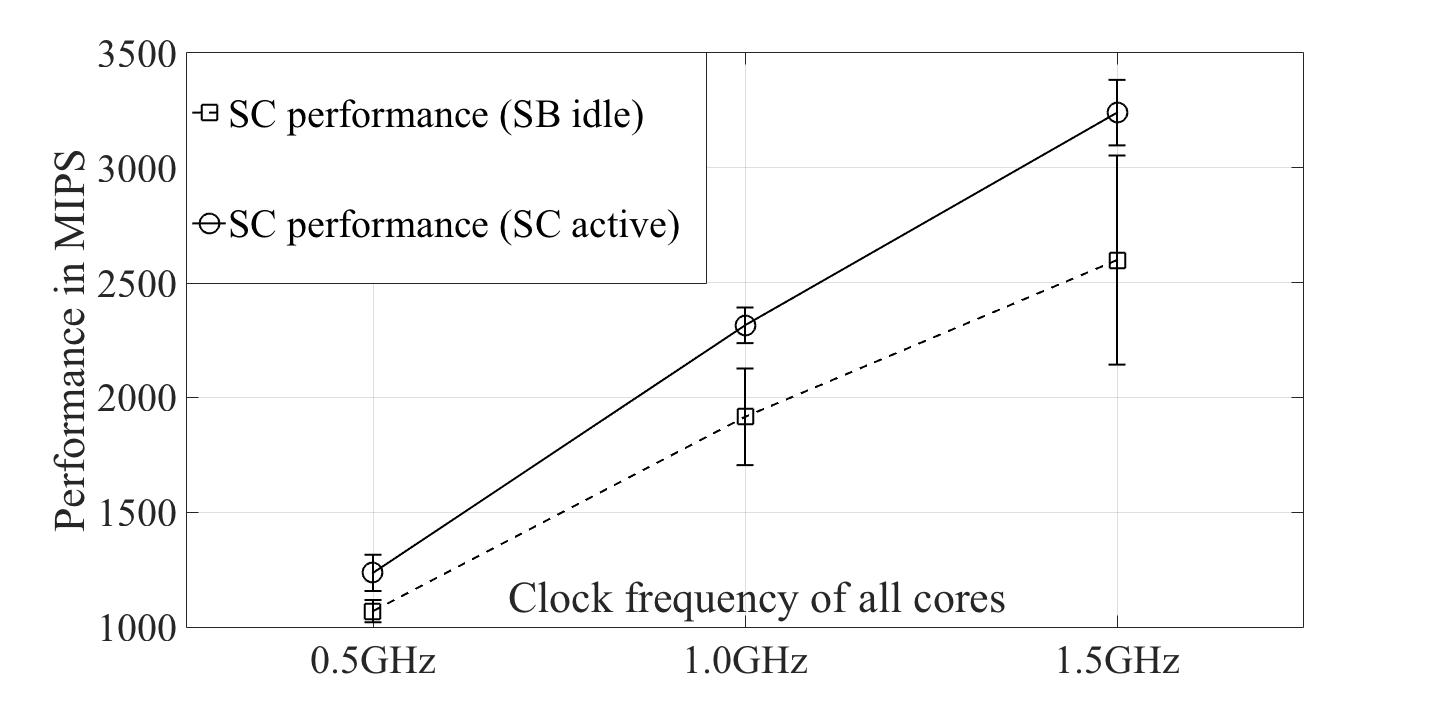}
		\caption{\textbf{Performance variation of the source core when source bank is `idle' and `active'}.}
		\label{fig:InterLayerPerformance}
	\end{subfigure}
	\caption{\textbf{Thermal coupling \textbf{Cases (b)} and \textbf{(c)}. The error bars are variances in temperature due to different ops/byte and physical locations of the source core}.}	
	\label{fig:CaseBC}
	\vspace{-4mm}
\end{figure}

\begin{figure}[!htb]
    \centering
    \begin{subfigure}[!htb]{0.45\textwidth}
		\includegraphics[width=3.5in,clip,trim={1.5cm 0 0 0}]{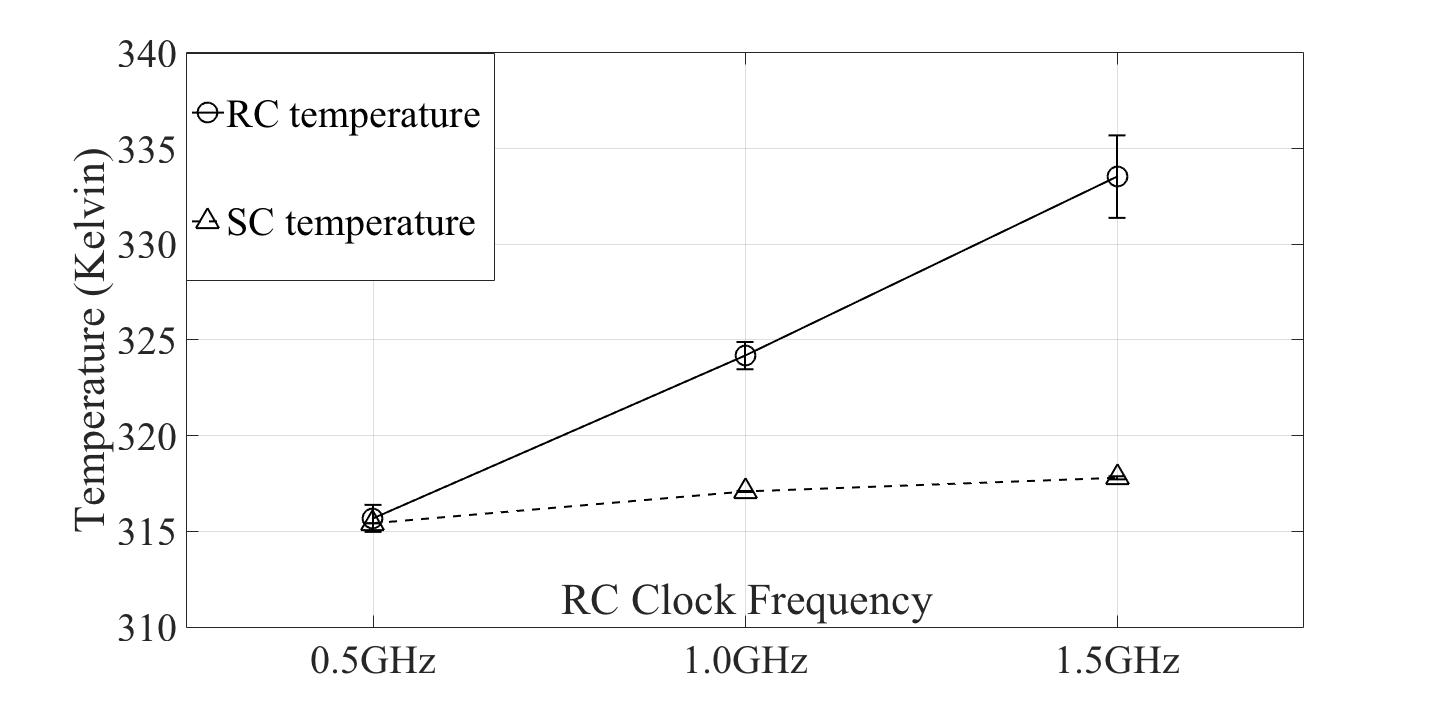}
		\caption{\textbf{Temperature variation of source core and remote core}.}
		\label{fig:IntraLayerTemperature}
	\end{subfigure}
	\vspace{-2mm}
	\begin{subfigure}[!htb]{0.45\textwidth}
		\includegraphics[width=3.5in,clip,trim={1.5cm 0 0 0}]{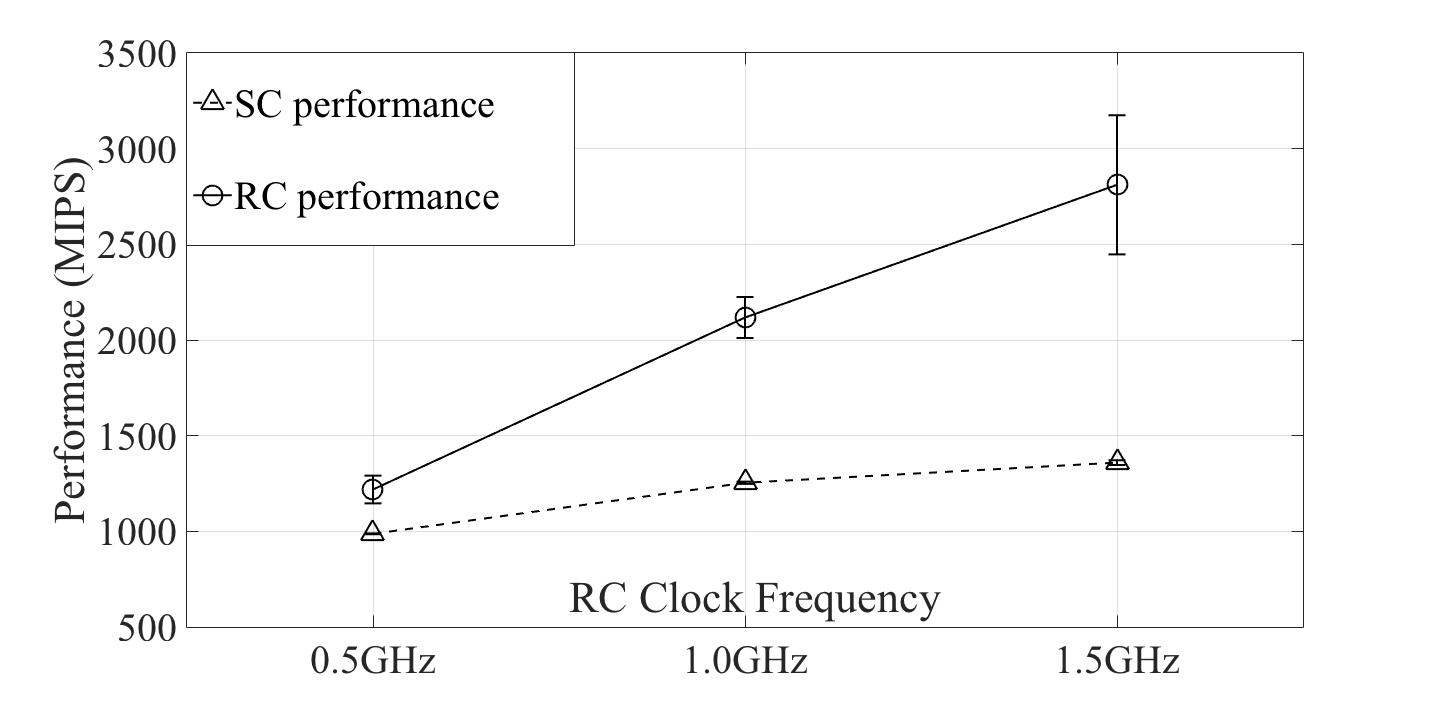}
		\caption{\textbf{Performance variation of source core and remote core}.}
		\label{fig:IntraLayerPerformance}
	\end{subfigure}
	\caption{\textbf{Thermal coupling \textbf{Case (d)}. The error bars are variances in temperature due to different ops/byte and physical locations of the source core}.}	
	\label{fig:CaseD}
	\vspace{-6mm}
\end{figure}

To completely understand the thermal coupling between the compute and memory layers, we divide this inter-layer thermal coupling into \textbf{Cases (b), (c)} and \textbf{(d)}. For \textbf{Cases (b)} and \textbf{(c)} we refer to Figure \ref{fig:CaseBC} and for \textbf{Case (d)} we refer to Figure \ref{fig:CaseD}. Before we present the analysis, it is essential to note that for the 3D architecture under consideration, in steady state, the core layer \textit{always} has the highest temperature when compared to upper layers. \\

\noindent \textbf{Case (b):} The heat flow between the SC and the SB is influenced by whether the SB is `active' or `idle'. The temperature trends for the SC and SB are presented in Figure \ref{fig:InterLayerTemperature}. When the SB is idle, the average SC temperatures are $312.7$, $319.3$ and $327.1$ Kelvin at $0.5$, $1.0$ and $1.5$GHz, respectively. But when the SB is active, the same SC temperatures increase by about $1$, $2$ and $4$ Kelvin for $0.5$, $1.0$ and $1.5$GHz, respectively. This clearly demonstrates the influence of memory addressing on the core layer temperatures. Not only does the average temperature rise with increase in frequency, but also the variance. At higher clock frequencies, thermal ramifications due to memory addressing patterns are more pronounced. The performance trend as seen if Fig. \ref{fig:InterLayerPerformance} is in accordance with our expectation, in that, instruction throughput benefits directly due to clock frequency increase.\\

\noindent \textbf{Case (c):} Moving along the same analysis path as before, for this case of thermal coupling, we wish to understand the variations in temperature of an `idle' core directly underneath an `active' L2 Cache bank. The temperature plots of the remote core (RC) and remote cache bank (RB) in Figure \ref{fig:InterLayerTemperature} illustrate this situation. Analogous to the previous case, bulk of the power dissipated by the idle RC underneath the active RB is on account of static power. Furthermore, as clock frequency increases, idle RC temperature can increase up to $5$ Kelvin higher than the lowest temperature on the core layer. \\

\noindent \textbf{Case (d):} This case is essentially a superposition of \textbf{Cases (b)} and \textbf{(c)}. The experiments here attempt to replicate a scenario where multiple cores can access a single L2 Cache bank. As described in Fig. \ref{fig:Nomenclature}, both the SC and the RC access the RB. Since RC is not idle anymore, we observe an increasing trend in its temperature with clock frequency. The slope of this increase however, is slightly steeper when compared to SC temperature (SB active) in Figure \ref{fig:InterLayerTemperature}. Furthermore, the increase in the clock frequency of the \{RC - RB\} voltage island causes the performance of RC and SC to improve (See Fig. \ref{fig:IntraLayerPerformance}). Due to difference in network delays however, slope of the performance curve for the SC is much smaller than that for the RC.

In summary, microbenchmark characterization of the 3D stack sheds light on subtle yet key insights. The EHC of an application thread is affected not only by its own phases but also by memory addressing patterns of neighboring cores. A greedy approach to maximizing performance can indeed utilize the thermal headroom of the package but may not deliver the best energy efficiency (ops/Joule). Consequently, the higher temperatures especially in thermally constrained environments such as the one under consideration, can increase thermal stresses and localized hotspots in turn reducing lifetime reliability of the device. Nevertheless, \textit{maximizing performance in the face of unavoidable thermal coupling, necessitates a strategy that cooperatively balances performance, energy and temperature}.

\section{TRINITY}
\label{sec:DTM}
In this section, we present our proposed approach, TRINITY, an online DVFS controller that dynamically balances the \underline{three} parameters: performance, energy and temperature to completely utilize the EHC in a \underline{3D stack}. TRINITY is, (i) application agnostic, (ii) self-tuning, (iii) distributed (per-core), (iv) based on numerical optimization, and (v) computationally inexpensive to implement. The TRINITY controllers on each core are implicitly coupled via temperature. Therefore, the individual actions taken by the network of controllers works towards making the best use of the EHC. We now present a detailed description of the system models, optimization problem and the solution approach.

\subsection{System Models}
\label{sec:SystemModel}
\begin{figure}
\centering
    \begin{subfigure}[!htb]{0.45\textwidth}
        \centering
    	\includegraphics[width=2.3in,height=1.2in]{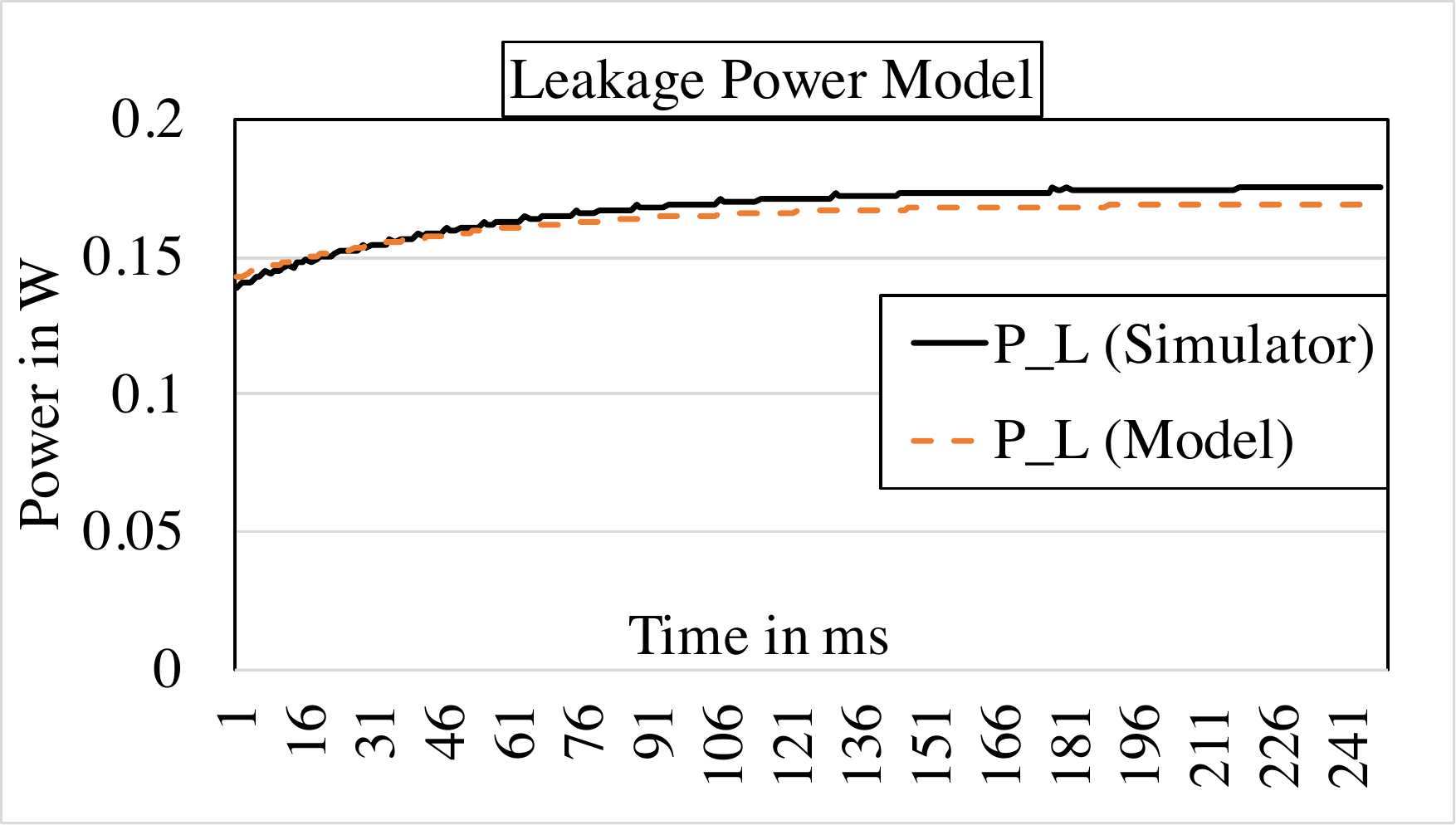}
    	\caption{\textbf{Leakage Power Simulator vs. Model}.}
    	\label{fig:LeakageModelValidation}
    \end{subfigure}
    \begin{subfigure}[!htb]{0.45\textwidth}
        \centering
    	\includegraphics[width=2.3in,height=1.2in]{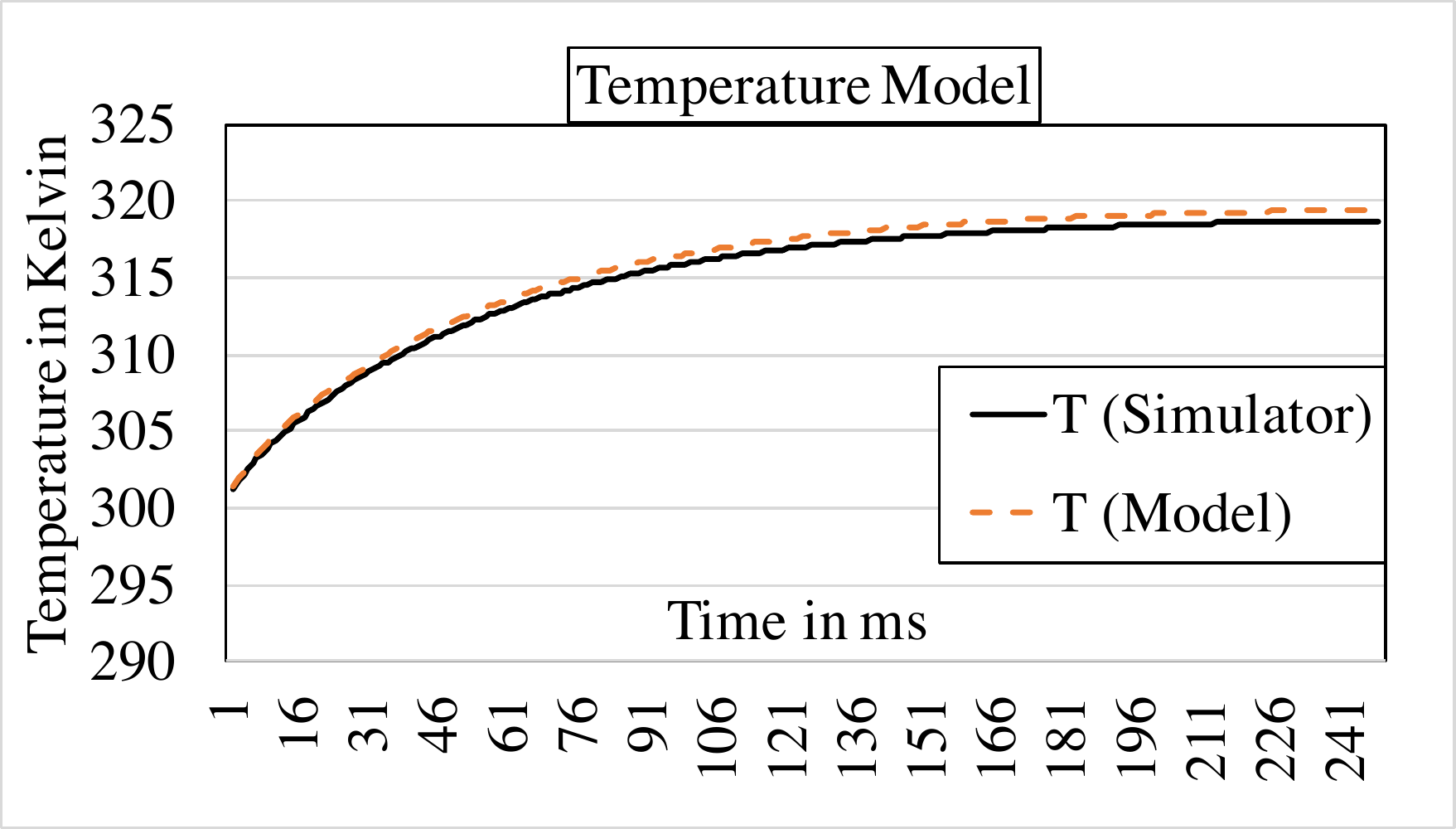}
    	\caption{\textbf{Temperature Model Simulator vs. Model}.}	
    	\label{fig:TemperatureModelValidation}
    	\vspace{-4mm}
    \end{subfigure}
    \caption{\textbf{Leakage power and temperature model}.}
    \vspace{-6mm}
\end{figure}

\textbf{\textit{Power Model}}: We linearize the power model in \cite{rao2015temperature} and arrive at a third order polynomial which captures both leakage and dynamic power of the core. The equation is as follows:
\begin{equation}
	P_k = \alpha f^3_k + \beta f_k + \gamma T_k + \delta f_k T_k + \epsilon
	\label{eqn:PowerModel}
\end{equation}	
where $k$ represents the sample time instant, $f_k$ and $T_k$ are clock frequency and core temperature, respectively. The first term models the dynamic power and the last four terms of the equation represent leakage power. Since leakage power is strongly correlated with the technology node and packaging parameters, via non-linear regression, we calculate $\beta, \gamma, \delta$ and $\epsilon$ offline (See Table \ref{tab:OfflineParams}). To enable TRINITY to be application agnostic, the constant $\alpha$, which represents the \textit{activity factor} is therefore determined online. Figure \ref{fig:LeakageModelValidation} shows that our approximation for the leakage power is within $\pm 5$mW of the value measured on the simulator.
\begin{table}[!htb]
	\centering
	\vspace{-2mm}
	\caption{\textbf{Parameters estimated offline}.}
	\label{tab:OfflineParams}
	\vspace{-2mm}
	\begin{tabular}{|c|c||c|c|}
		\hline
		$\beta$ 		& -426.7$\times 10^{-3}$ & $a_1$ & 0.9998\\ \hline
		$\gamma$ 	& 0.674$\times 10^{-3}$		& $b_1$ & 8.46\\ \hline
		$\delta$ 		& 1.618$\times 10^{-3}$		& $c_1$ & 37\\ \hline
		$\epsilon$ 	  & -90.38$\times 10^{-3}$	 &  $\Delta t$ & $1$ms \\ \hline
	\end{tabular}
	\vspace{-4mm}
\end{table}

\textbf{\textit{Temperature Model}}: Temperature at any given point in the 3D stack at any given time $t$ is given by the dynamical equation
\begin{equation}
\dot{\textbf{T}}(t) = \textbf{AT}(t) + \textbf{BP}(t)
\label{eqn:FullTemperatureModel}
\end{equation} 
where $\textbf{T}(t), \textbf{P}(t) \in \mathds{R}^M$ are the temperature and power vectors, respectively and the matrices $\textbf{A}$ and $\textbf{B}$ consist of the thermal resistance and capacitance of the 3D stack \cite{han2007tilts}. In each of the 6 layers in the 3D stack, broadly, we have 16 power dissipating elements, therefore, $M = 16 \times 6 = 96$. The large $\textbf{A}$ and $\textbf{B}$ matrices capture the inter-layer and intra-layer thermal coupling allowing for an accurate estimation of the temperature trajectory. At this juncture it is relevant to note that for the 3D stack under consideration, we observe a time constant for the rise in temperature of approximately $40$ms and thus the settling time is around $200$ms. These numbers are in agreement with practical observations \cite{paul2013cooperative}. Solving an optimization problem as described in the Introduction becomes increasingly computationally intensive as the dimensionality of the model increases (typically $O(M^3)$). Instead of using Eqn. \ref{eqn:FullTemperatureModel} we make an observation that discretizing and linearizing Eqn. \ref{eqn:FullTemperatureModel} for a short duration of time $\Delta t$, reduces the model complexity significantly from $O(96^3) \rightarrow O(1)$. The price paid for this reduction in complexity is in the ability to accurately predict future temperature. Nonetheless, the temperature of a core can now be estimated $\Delta t$ seconds into the future using the following scalar equation:
\begin{equation}
T_{k+1} = a_1 T_k + \Delta t (b_1 P_k + c_1)
\label{eqn:SimpleTemperatureModel}
\end{equation}
where we observe that up to $\Delta t = 1$ms, the simplified temperature model is within $1$ Kelvin as compared with values obtained from the simulator (See Fig. \ref{fig:TemperatureModelValidation}). Analogous to the power model, the constants $a_1, b_1$ and $c_1$ are dependent on the technology node and packaging design choices. Therefore they are estimated offline via non-linear regression (See Table \ref{tab:OfflineParams}). The temperature estimate $T_{k+1}$ depends on the \textit{measured} values at time sample $k$ and thus does not accumulate modeling errors at each time step. 

\textbf{\textit{Performance Model}}: We choose instruction throughput i.e. MIPS as the metric. Performance is related to the clock frequency of the core via the following equation:
\begin{equation}
\chi_k(u_k) = IPC_k \cdot f_k
\label{eqn:PerfModel}
\end{equation}
where $IPC_k$ is Instructions-Per-Cycle and $f_k$ is the core clock frequency at sample time $k$. As shall be described shortly, this linear approximation counterbalances temperature rise and is therefore sufficient for the purposes of the optimization problem under consideration.

\subsection{Solution Strategy}
\label{sec:Optimization}
The objective of the problem we wish to solve is encoded mathematically as follows: 
\begin{align}
\label{eqn:Objective}
\max_{f_{k+1}} \quad & Q z^2_{k+1} + R \chi_k(f_{k+1})\\
\label{eqn:ControlBounds}
\text{subj. to} \quad \nonumber\\
\underline{f} \leq & f_{k+1} \leq \bar{f} \\
\label{eqn:StateBounds}
0 < & z_{k+1}
\end{align}
where $f_{k+1}$ is the core frequency which is within the bounds $\underline{f} = 0.5$GHz and $\bar{f} = 1.5$GHz. The term $z_{k+1} = T_{MAX} - T_{k+1}$, where $T_{k+1}$ is the temperature of the core in Kelvin (Eqn. \ref{eqn:SimpleTemperatureModel}) and $T_{MAX}$ is an upper bound for a core's temperature which we set to $355 \text{Kelvin} = 85^0C$. The cost function described by Eqn. \ref{eqn:Objective} consists of two parts, the former that penalizes increase in temperature and the latter that rewards performance. The weights $Q$ and $R$ ($Q,R > 0$  for problem feasibility) are tuning parameters that can be modified at run-time to give variable importance to performance and temperature. We set $Q = 1$ and allow $R$ to \textit{tune itself} at run-time. Equations \ref{eqn:Objective} - \ref{eqn:StateBounds} are solved periodically after every $\mathds{T}$ seconds by each core \textit{independently} to determine $f^*_{k+1}$, the clock frequency that maximizes the cost.

\begin{figure}[!htb]
    \centering
    \vspace{-4mm}
    \includegraphics[width=3.5in,height=1.2in,clip,trim={1.5cm 3cm 0 4cm}]{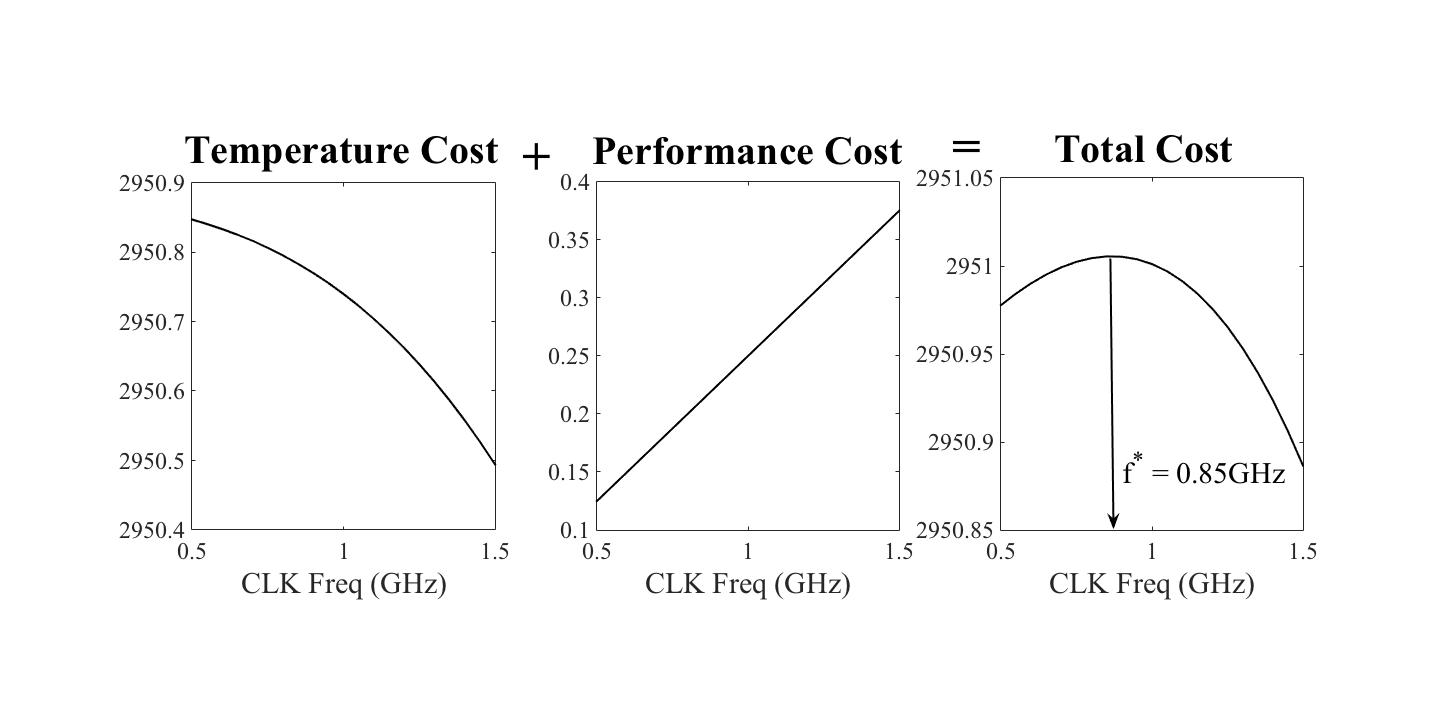}
    \vspace{-6mm}
    \caption{\textbf{Behavior of the optimization cost}.}
    \label{fig:BalancingFigure}
    \vspace{-4mm}
\end{figure}

\begin{figure}[!htb]
\centering
\vspace{-2mm}
	\includegraphics[width=0.2\textwidth,height=2in]{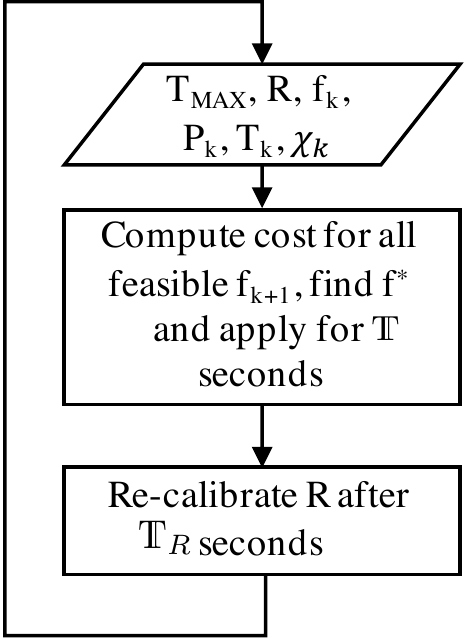}
	\vspace{-2mm}
	\caption{\textbf{TRINITY Algorithm}.}
	\label{fig:Algorithm}
	\vspace{-2mm}
\end{figure}

The intuition behind the problem definition is as follows: Consider an application whose performance saturates beyond a particular clock frequency and does not vary with time. The periodic calculation of $f^*_{k+1}$ drives the system eventually towards a point where the temperature of the core reaches steady state. This steady state temperature is nothing but the EMT and any further increase in the clock frequency will reduce the cost thereby satisfying our original goal of maximally utilizing the EHC. For a particular choice of $R$, the behavior of the objective function is illustrated in Figure \ref{fig:BalancingFigure}. The value $\mathds{T}$, referred to as the control cycle, is a design parameter which has to be at least greater than (i) measurement, (ii) actuation, and (iii) computation delays. On processors available in the market currently, measurement and actuation delays are approximately $10$s of micro seconds \cite{mazouz2014evaluation}. The control cycle also depends on the model accuracy since, as observed in the previous section, the simplified temperature model has sufficient accuracy up to a duration of $1$ms. Therefore, we choose $\mathds{T} = 1$ms in our experiments. We assign 21 clock frequencies between $0.5 - 1.5$GHz spaced $50$MHz apart. To solve the problem described in Eqn. \ref{eqn:Objective}, the three steps of the algorithm are listed in Fig. \ref{fig:Algorithm}. Since each core solves the optimization problem independently, computing $f^*_{k+1}$ requires finding the maximum element in an 21 length array. 

\begin{figure}[!htb]
\centering
\vspace{-4mm}
	\includegraphics[width=0.4\textwidth,height=3in,clip,trim={0 1cm 0 0}]{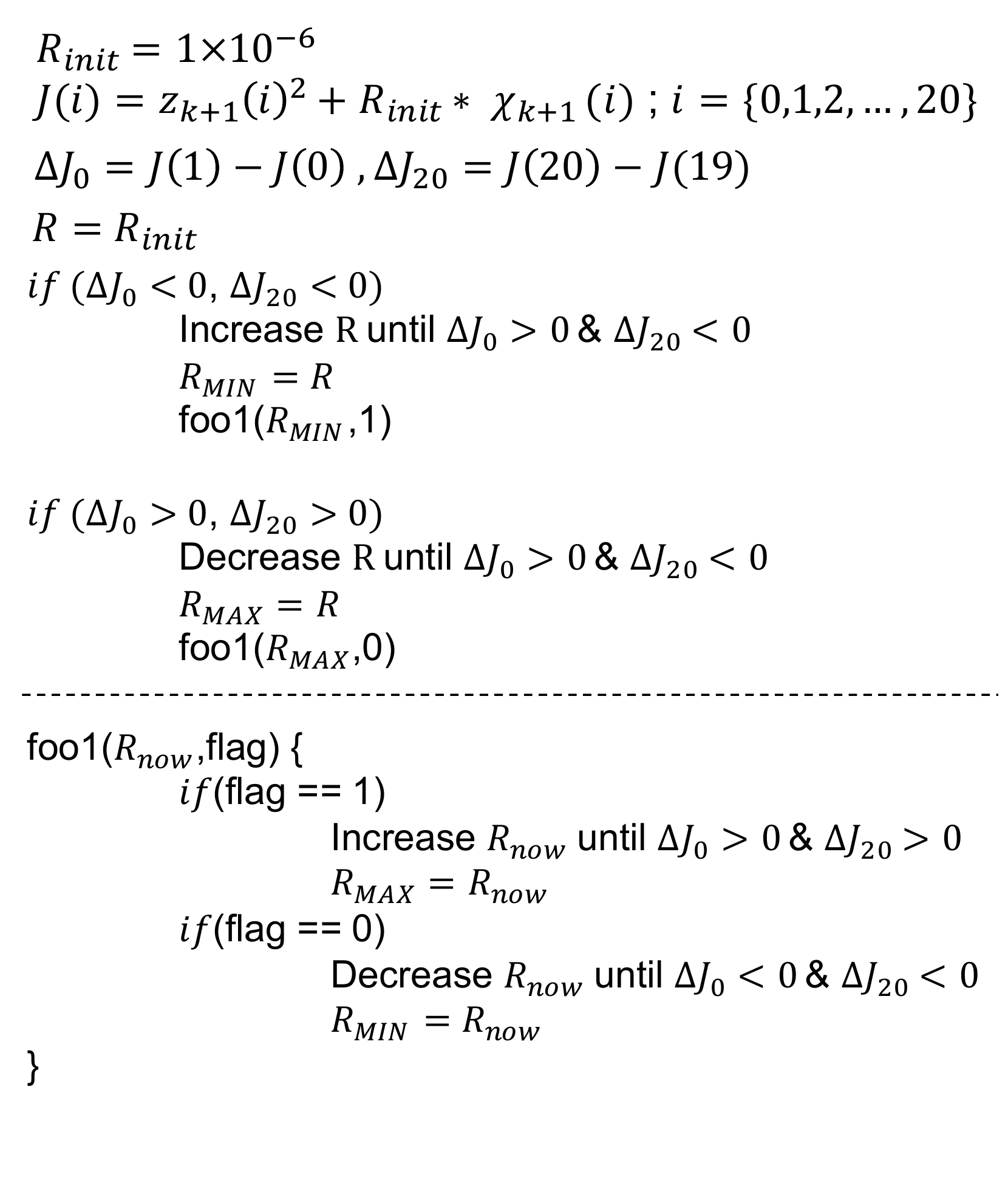}
	\vspace{-4mm}
	\caption{\textbf{Pseudo code for re-calibrating $R$}.}	
	\label{fig:RAlgorithm}
	\vspace{-6mm}
\end{figure}

The tuning parameter $R$ influences all three variables: temperature, performance and leakage energy. In fact, $R \in [R_{min},  R_{max}]$ such that for $R < R_{min}$, $f^*_{k+1} = \underline{f}$ and for $R > R_{max}$, $f^*_{k+1} = \bar{f}$. Emphasizing temperature over performance leads to lower leakage energy whereas greater importance to performance could potentially lead to wasted leakage energy. Therefore it is essential to choose an appropriate value in order to exact the behavior desired. Fixing the value of $R$ is one approach. However, we observe that such a strategy, (i) makes the solution application specific (ii) requires extensive time consuming offline analysis, and (iii) could easily push the controller into saturation where $f^*_{k+1}$ will be remain at either $\underline{f}$ or $\bar{f}$ for prolonged periods of time. In order to adapt to dynamically varying application phases, we allow $R$ to re-calibrate itself periodically. We call this period $\mathds{T}_R \geq \mathds{T}$. The pseudo code for the re-calibration is described in Figure \ref{fig:RAlgorithm}. The re-calibration step basically determines the bounds for $R$ i.e. $[R_{min}, R_{max}]$ and calculates the next value as $R = R_{min} + \eta (R_{max} - R_{min})$ where $\eta \in [0,1]$. For the OoO core under consideration, we set $\eta = \sqrt{IPC_k / IPC_{max}}$ with $IPC_{max} =$ Issue Width $= 4$. The IPC ratio heuristic is a means to obtain information about the compute or memory boundedness of the application. The square root of the ratio is chosen to push $R$ towards $R_{max}$ (and thus better performance).


\section{Results}
\label{sec:Results}
In this section we evaluate TRINITY. We first describe the simulation environment and list the benchmark applications used. Next we present an evaluation of the proposed control scheme in detail. We implement a DVFS strategy similar to the \texttt{ondemand} Linux CPU governor on the simulator and compare it against TRINITY. We also present results by fixing the core frequencies to $0.5$, $1.0$ and $1.5$GHz. Since TRINITY attempts to balance performance, energy and temperature, we use Energy Delay$^2$ Product (ED2P) along with temperature as the primary comparison metric. In what follows, we also use Energy Delay Product (EDP), Energy Efficiency (ops/Joule), performance (MIPS) and lifetime reliability measured as Mean Time To Failure (MTTF) to understand the implications of using TRINITY.

\subsection{Experimental Framework}
\label{sec:Framework}
\begin{figure}[!htb]
	\centering
    \includegraphics[width=0.45\textwidth]{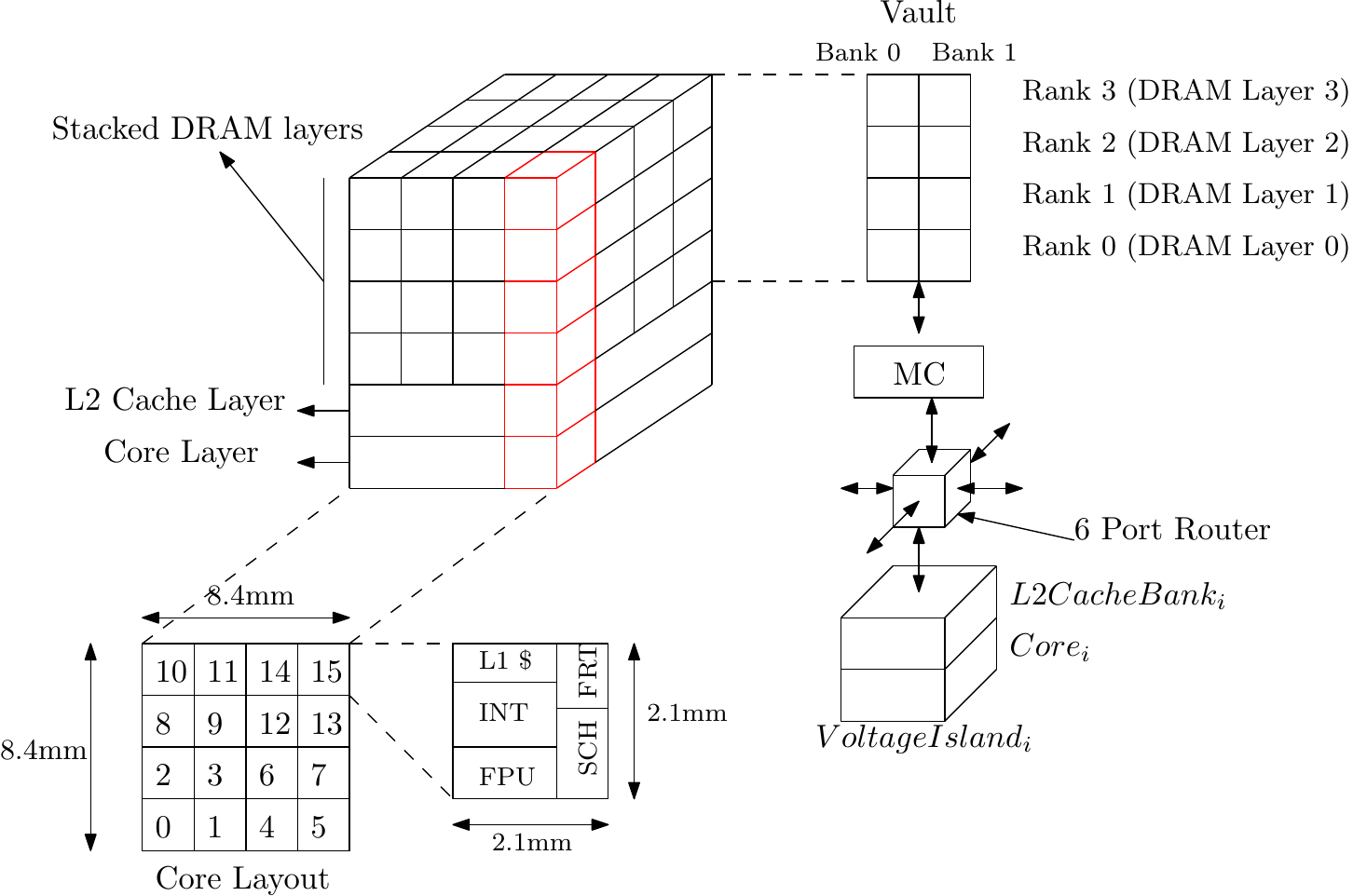}
    \vspace{-2mm}
    \caption{\textbf{Physical layout of the 3D stacked architecture}.}
    \label{fig:SystemFunctional}
    \vspace{-4mm}
\end{figure}

\begin{figure}[!htb]
    \centering
    \includegraphics[height=1.2in,width=1.4in]{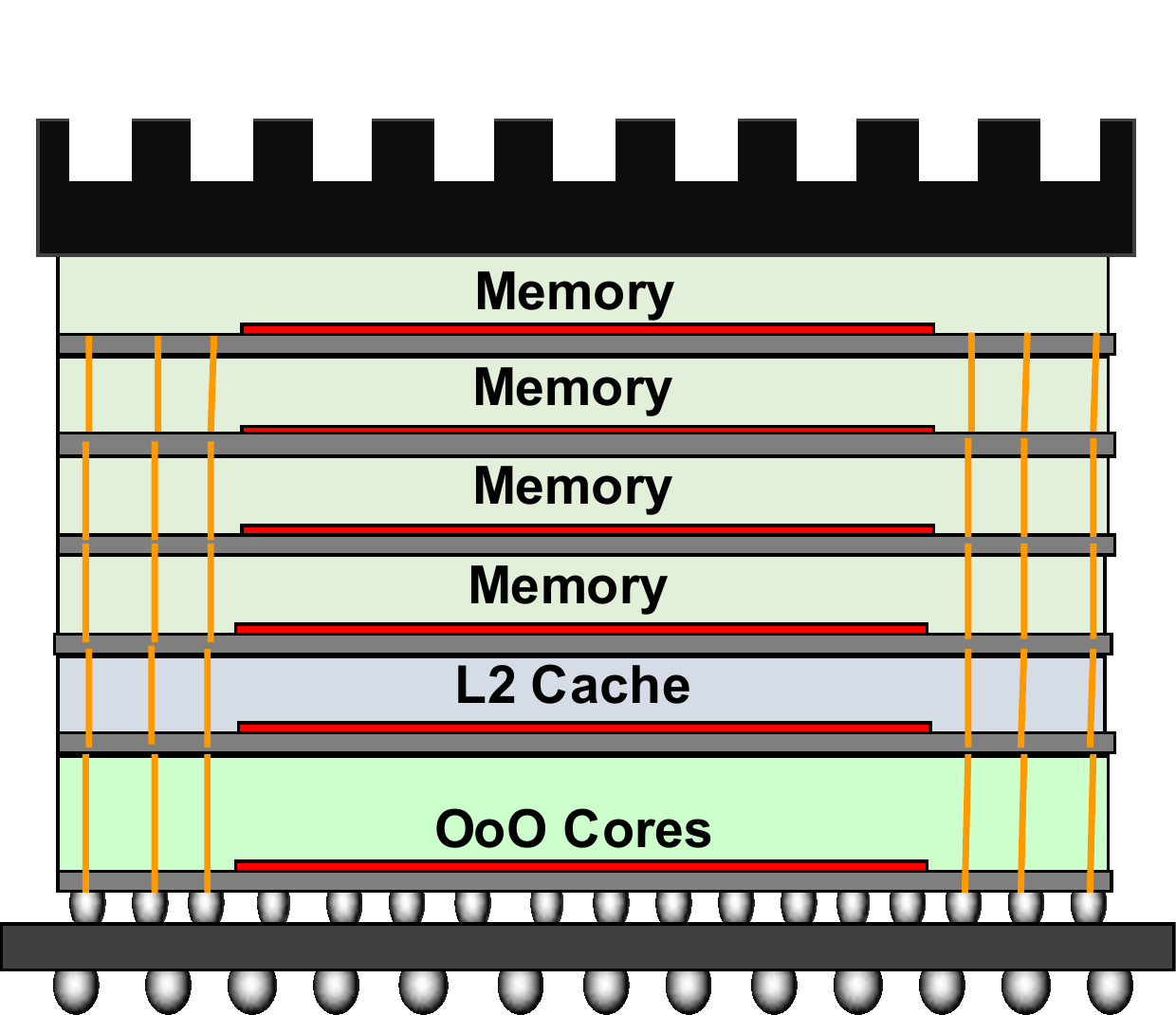}
    \vspace{-2mm}
    \caption{\textbf{Functional description of the 3D stack}.}
    \label{fig:SystemPhysicalLayout}
    \vspace{-6mm}
\end{figure}

The physical layout is shown in Fig. \ref{fig:SystemFunctional} with the dimensions listed in Table \ref{tab:SystemDescription} and Figure \ref{fig:SystemPhysicalLayout} represents the functional diagram of the 3D stacked architecture. The 3D stacked architecture consists of 16 Out-of-Order (OoO) cores with two levels of cache hierarchy \cite{beu2011manager}, interfacing an HMC style \cite{hmcconsortium} 4GB DRAM via an interconnection network. The simulator is also equipped with power and thermal estimation framework called Kitfox \cite{song_tcpmt2015}. Kitfox internally models power consumption based on McPat \cite{li2009mcpat} and the thermal calculations are done using 3D-ICE \cite{sridhar20103d}, both scaled to $16$nm. The front-end for the cycle level simulator \cite{wang2014manifold} is a multicore emulator called Qsim \cite{kersey2012universal} that boots a Linux kernel and executes applications of interest. The x86 instruction streams thus generated are fed into the OoO core timing model. We use DRAMSim2 \cite{rosenfeld2011dramsim2} as the DRAM timing simulator whose voltage and timing numbers are modified based on the work in \cite{hassan2016understanding}. 
\begin{table}[!htb]
	\centering
	\vspace{-2mm}
	\caption{\textbf{Simulation framework parameters. Technology node is 16nm}.}
	\label{tab:SystemDescription}
	\vspace{-2mm}
	\begin{tabular}{|p{2.1cm}||p{5.5cm}|}
	    \hline
		\textbf{Component}          &   \textbf{Parameters and Values} \\ \hline\hline
		Processor                   &   Out-of-Order, 6-stage pipeline, 4-wide issue/commit, $0.5 - 1.5$GHz \\ \hline
		L1 Cache per core (16KB)	&	Private, 8-way, LRU replacement, 32 MSHRs, 64B lines, 1-cyc hit \& lookup time \\ \hline
		L2 Cache per bank (2MB)	    &	16 banks in total, shared, 8-way, LRU replacement, 128 MSHRs, 64B lines, 24-cyc hit \& lookup time \\ \hline
		Network (1GHz)              &	$4\times4$ torus ring, 6 port router, baseline x-y routing \\ \hline
		Memory Controller   	    &	16 MCs in total, rank then bank round robin, close page, Addr-map- chan:row:bank:rank:col \\ \hline
		DRAM config per vault	    &	256MB, 1-channel, 4 ranks, 2 banks per rank, 64 bit bus @ 1600MHz \\ \hline \hline
        Heat Sink					&	Conventional heat sink, Heat transfer co-eff = $2.8\times10^{-8}W/\mu m^2K$ \\ \hline
        \multirow{3}{*}{Per-Layer}  &   TOP LAYER = BEOL: $25\mu$m \\
                                    &   SOURCE LAYER = SILICON: $10\mu$m \\
                                    &   BOTTOM LAYER = SILICON: $25\mu$m \\ \hline
	\end{tabular}
    \vspace{-4mm}
\end{table}

Various microarchitectural design choices for placement of cores, cache and DRAM exist as described in \cite{puttaswamy2007thermal}. The architecture described in this work is similar to the one proposed in \cite{li2006design}. We do however note that our microarchitectural design choice does not restrict the scope of the problem definition. The thermal and performance characterization insights can be extended to other 3D architectures as well.

\subsection{Benchmarks}
\label{sec:Benches}
We evaluate the optimization technique proposed over 6 benchmark applications from the PARSEC, SPLASH2x \cite{bienia2008parsec} and GraphBig \cite{nai2015graphbig} suite. Specifically, we choose \texttt{blackscholes}, \texttt{streamcluster}, \texttt{barnes} (PARSEC and SPLASH2x) and \texttt{kcore}, \texttt{pagerank} and \texttt{tc} (GraphBig). Each of the benchmark applications are executed with 16 threads. The GraphBig applications stress the memory whereas PARSEC and SPLASH2x stress the compute units thus giving a range of application behavior. 

\subsection{Analyzing TRINITY Performance}
\label{subsec:TRINITYPerf1}
\begin{figure*}[!htb]
    \centering
    \includegraphics[width=\textwidth,height=2in]{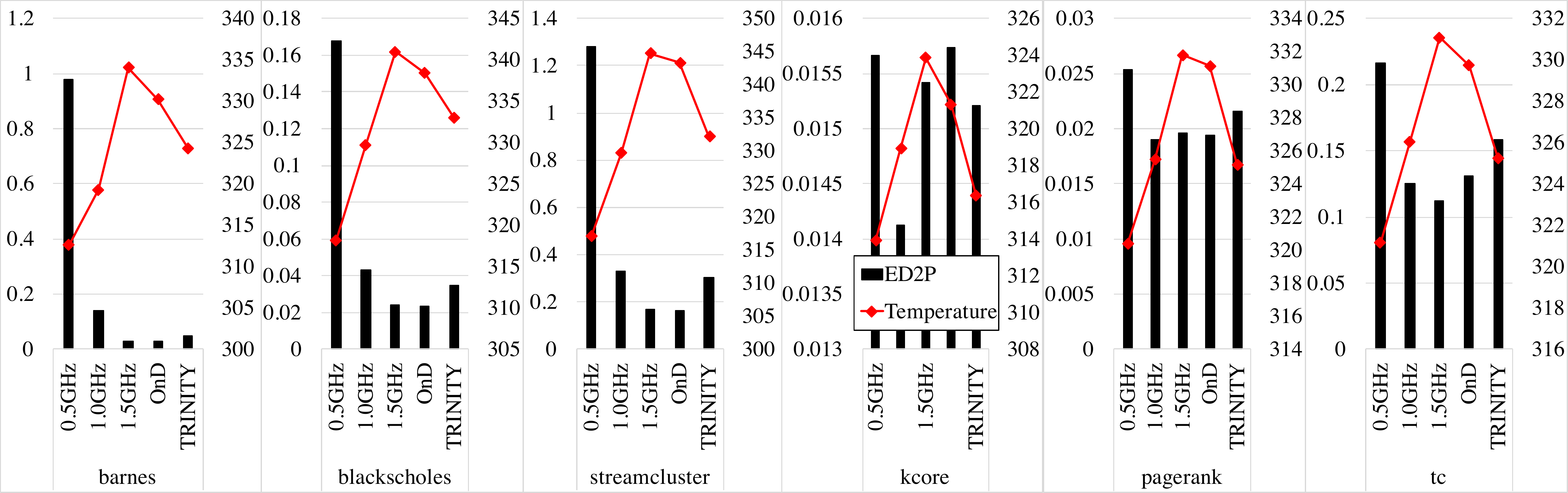}
    \vspace{-4mm}
    \caption{\textbf{Controller performance compared against fixed frequencies and \texttt{ondemand}. Controller Parameters: $T=1$ms and $T_R=5$ms. Left y-axis and right y-axis units are ED2P and Kelvin, respectively}.}
    \label{fig:Run1PerfTempComparison}
    \vspace{-4mm}
\end{figure*}

\begin{figure*}[!htb]
    \centering
    \includegraphics[width=\textwidth,height=2in]{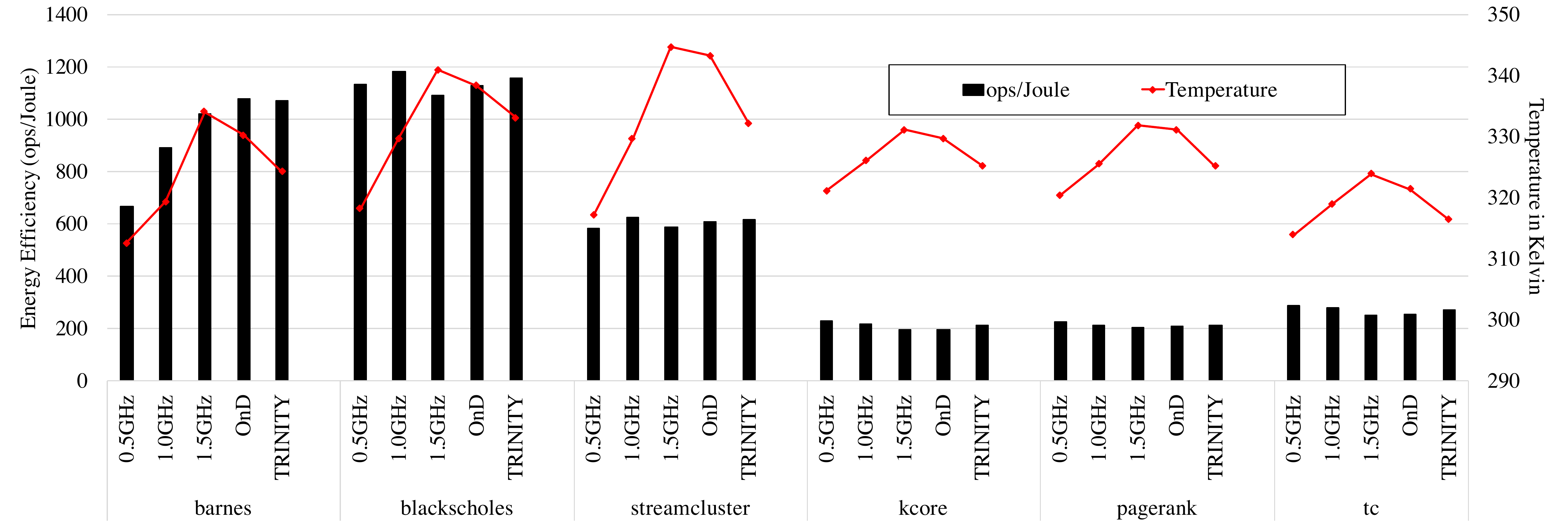}
    \vspace{-4mm}
    \caption{\textbf{Controller performance compared against fixed frequencies and \texttt{ondemand}. Controller Parameters: $T=1$ms and $T_R=5$ms. Left y-axis and right y-axis units are ops/Joule and Kelvin, respectively}.}
    \label{fig:Run1opsJouleTempComparison}
    \vspace{-4mm}
\end{figure*}

Energy efficiency and ED2P results are plotted in Figures \ref{fig:Run1opsJouleTempComparison} and  \ref{fig:Run1PerfTempComparison}, respectively, comparing TRINITY against the three fixed frequencies and \texttt{ondemand}. The right y-axis in both the figures represents the spatial average temperature of the core layer. The control cycle $\mathds{T}$ is set to $1$ms. Re-calibrating $R$ every control cycle increases the amount of computations performed by the controller and hence we set $\mathds{T}_R=5$ms.

The trend of ED2P is not the same for every benchmark. Consequently, the strategy to balance performance, energy and temperature (PET) should be different. In general, for compute intensive benchmarks (\texttt{blackscholes}, \texttt{barnes} and \texttt{streamcluster}), the highest clock frequency delivers the best performance and also results in the highest temperature. TRINITY on the other hand, trades performance for benefits in temperature (lower by $8$ Kelvin w.r.t $1.5$GHz and $6$ Kelvin w.r.t \texttt{ondemand}). In fact, the reduction in ED2P for the compute intensive benchmarks is due to the reduction in performance. Energy efficiency results however reveal that TRINITY and \texttt{ondemand} perform equivalently (< $5\%$ difference). We can therefore conclude that in the process of balancing PET for compute intensive benchmarks, TRINITY achieves similar energy efficiency as \texttt{ondemand} but keeps the temperature $6$ Kelvin lower.

Analyzing memory intensive benchmarks (\texttt{kcore}, \texttt{pagerank} and \texttt{tc}), there is no appreciable improvement in performance (MIPS). For example, average MIPS for \texttt{kcore} at $0.5$, $1.0$ and $1.5$GHz is $4360.3$, $5074.2$ and $5265.2$, respectively. Possessing \textit{apriori} knowledge that the application to be executed is memory intensive, could lead to choosing the lowest clock frequency as a possible strategy. While it certainly keeps the entire 3D stack at a lower temperature, ED2P suffers significantly. In these situations, TRINITY tunes $R$ in such a way that the lower half of the clock frequencies ($0.5$ - $1.0$GHz) are chosen in the memory intensive phases. For all three benchmarks, temperature of the core layer is at most as high as $1.0$GHz. Performance and ED2P, while certainly better than $0.5$GHz, are $5\%$ and $13\%$ worse than $1.0$GHz, respectively. On closer analysis of the controller data, we observe that predicting performance for the upcoming control cycle based on the previous control cycle, sometimes causes TRINITY to choose a clock frequency that does not maximize performance for the EHC. These mispredictions are mitigated to a certain extent by re-calibrating $R$ every $\mathds{T}_R$ seconds but it is one of the limitations of TRINITY.

Energy efficiency, similar to ED2P, has different trends for different benchmarks as shown in Figure \ref{fig:Run1opsJouleTempComparison}. While it increases for \texttt{barnes} as clock frequency is increased, it reduces for \texttt{kcore}. A detailed breakdown of different comparison metrics is shown in Figure \ref{fig:Run1ParsecGraphSeparateComparison}. Note that trading off performance (MIPS), affects EDP and ED2P directly. Energy efficiency of TRINITY however, is very similar to \texttt{ondemand} (3\% better). This is on account of reduced temperature. 


\begin{figure*}[!htb]
    \centering
    \includegraphics[width=0.9\textwidth,height=2.2in]{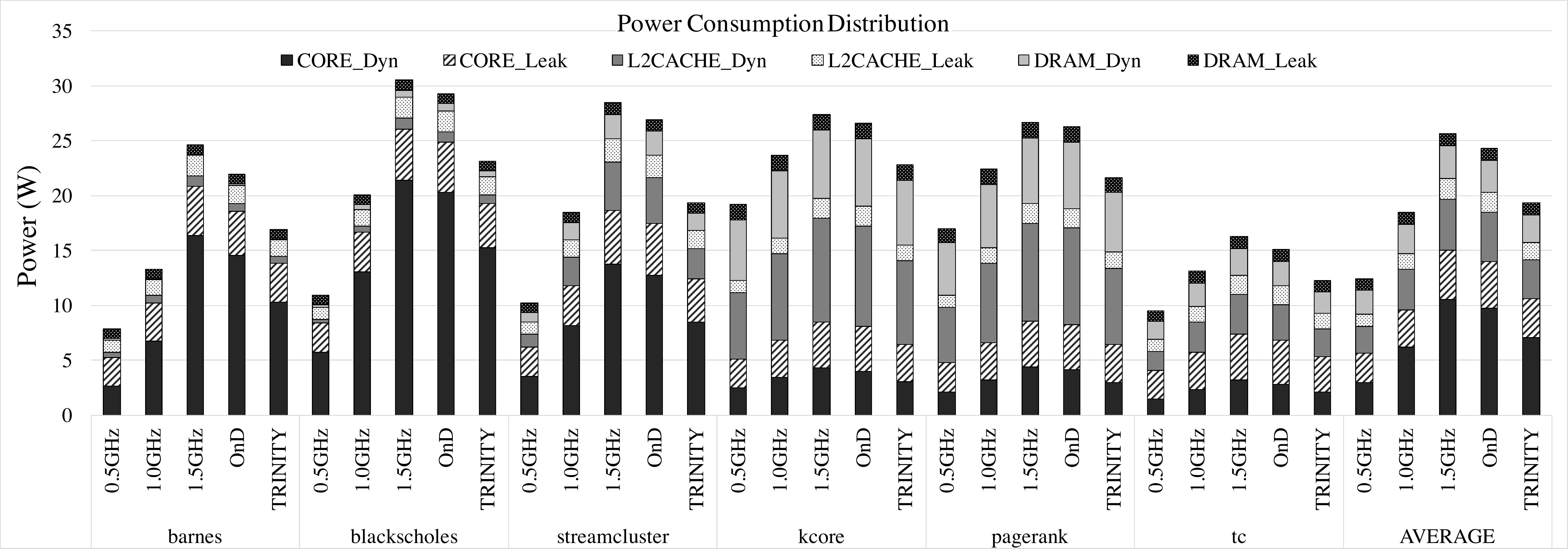}
    \vspace{-2mm}
    \caption{\textbf{Average power consumption by TRINITY compared against fixed frequencies and \texttt{ondemand}. Controller Parameters OPT1: $T=1$ms and $T_R=5$ms}.}
    \label{fig:Run1PowerComparison}
    \vspace{-6mm}
\end{figure*}

\begin{figure*}[!htb]
    \centering
    \includegraphics[width=\textwidth,height=2in]{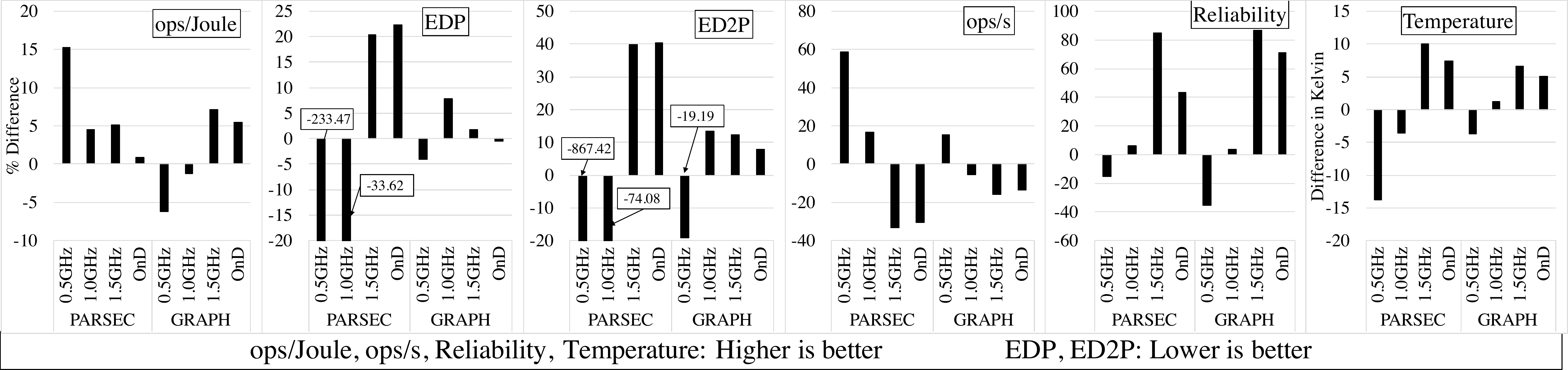}
    \vspace{-2mm}
    \caption{\textbf{TRINITY compared against fixed frequencies and \texttt{ondemand}. Controller Parameters OPT1: $T=1$ms and $T_R=5$ms}.}
    \label{fig:Run1ParsecGraphSeparateComparison}
\end{figure*}

To understand the source of temperature reduction, we plot the average power dissipated at individual layers in Figure \ref{fig:Run1PowerComparison}. The x-axis represents different DVFS options for each benchmark and the y-axis shows average power in Watts. Total power for each DVFS setting is broken down into dynamic and leakage power for the core, L2 Cache and DRAM layers. This distribution of power helps understand the primary source of power consumption for each benchmark application. \texttt{blackscholes} and \texttt{barnes}, both compute intensive, consume majority of the power in the core layer. \texttt{kcore} and \texttt{pagerank} being memory intensive consume greater power in the L2 Cache layer, specifically dynamic power in L2 Cache. Additionally, DRAM dynamic power is higher as well due to increased L2 Cache misses. \texttt{streamcluster}, unlike \texttt{blackscholes} and \texttt{barnes} shows increasing L2 Cache power for increasing frequencies. Finally, although \texttt{tc} consumes approximately same amount of power in both core and cache layers, analyzing average MIPS for the three fixed frequencies and reveals that \texttt{tc} is indeed memory intensive ($2741.3$, $3633.8$ and $4026.7$ MIPS at $0.5$, $1.0$ and $1.5$GHz, respectively).

As seen in Fig. \ref{fig:Run1PowerComparison}, the bulk of the power reduction (consequently temperature) comes from reducing dynamic power consumption of the core and cache layers. This is intuitive since DVFS implemented by TRINITY directly affects only the core and the corresponding L2 Cache bank. As compared to \texttt{ondemand}, dynamic power of the core and cache layers reduce by $38.5\%$ and $28.2\%$, respectively. Furthermore, with respect to \texttt{ondemand}, TRINITY is also able to reduce leakage power of the core and cache layers by $18\%$ each. We attribute the power reduction to the on-line adaptation of $R$. In memory intensive parts of the application, $\eta$ is low ($ < 1$) thus guiding the controller to choose the lower end of the clock frequencies. In compute intensive regions, $\eta$ is high ($ > 2$) allowing for higher clock frequencies to be chosen. 
\subsection{Impact on Lifetime Reliability}
\label{sec:Reliability}
Changes in operating temperatures and voltages incur significant impacts on reliability.
Two dominant reliability mechanisms, electromigration (EM) and time-dependent dielectric breakdown (TDDB), are used to evaluate the reliability implications of TRINITY, compared to that of other execution modes.
We used the reliability models and parameters from the work in \cite{song2015managing} and references therein.
Equations \ref{eqn:EM} and \ref{eqn:TDDB} show the reliability models of EM and TDDB, expressed as mean-time-to-failure (MTTF).
\begin{equation}
\text{MTTF}_{\text{EM}}=
A_{\text{EM}} \times \frac{1}{t_{\text{act}}} \times V^{-n} \times e^{\frac{E_{a}}{kT}}
\label{eqn:EM}
\end{equation}
\vspace*{-15pt}\leavevmode
\begin{equation}
\text{MTTF}_{\text{TDDB}}=
A_{\text{TDDB}} \times \frac{1}{t_{\text{act}}} \times V^{-c(a + bT)} \times e^{\frac{x+y/T+zT}{kT}}
\label{eqn:TDDB}
\end{equation}
In the reliability equations, $V$ and $T$ are operating voltage and temperature.
$t_{\text{act}}$ ($0 \le t_{\text{act}} \le 1$) is active-state residency obtained from the execution time of each workload.
For instance, $t_{\text{act}} = 0.5$ means that a workload utilizes the computing system for 50\% of time.
It is assumed that the system can be ideally power-gated for the remaining period and thus has no reliability impact;
the system may be used to process other workloads, but  resulting reliability impacts contribute to those workloads.
$k$ is Boltzmann's constant, and other parameters are model-dependent scaling parameters \cite{song2015managing}.
As shown in Eqn. \ref{eqn:EM}, EM is primarily accelerated by temperature, and voltage has a secondary effect.
In fact, a few degrees of average temperature change throughout the lifetime can easily produce several months to years of EM variations.
On the other hand, TDDB is more sensitive to voltage changes, but temperature also has a non-negligible effect.
The results show that TRINITY achieves 59\%, 88\%, and 6\% better reliability than the \texttt{ondemand} and two fixed frequency executions at 1.5GHz and 1.0GHz, respectively.
However, it trades performance improvement with 24\% reduction in reliability when compared with the 0.5GHz fixed frequency execution.

\subsection{Effect of TRINITY Parameter Variations}
As mentioned in the previous sections, one of our goals with TRINITY is to design it with practical implementation in mind. Simplifying the model and reducing computational complexity reduces the number of parameters that can be manually tuned. We study the sensitivity of TRINITY to variations in $\mathds{T}$ and $\mathds{T}_R$, which are the only manually tuned parameters. Reducing the control cycle duration and $\mathds{T}_R$ has the benefit of capturing rapidly varying application phases. But it also increases the amount of controller computations per unit time. We compare three different cases: (1) OPT1 ($\mathds{T}=1$ms, $\mathds{T}_R=5$ms), (2) OPT2 ($\mathds{T}=1$ms, $\mathds{T}_R=10$ms), and (3) OPT3 ($\mathds{T}=0.5$ms, $\mathds{T}_R=5$ms).  

\begin{figure*}[!htb]
    \centering
    \includegraphics[width=\textwidth,height=2in]{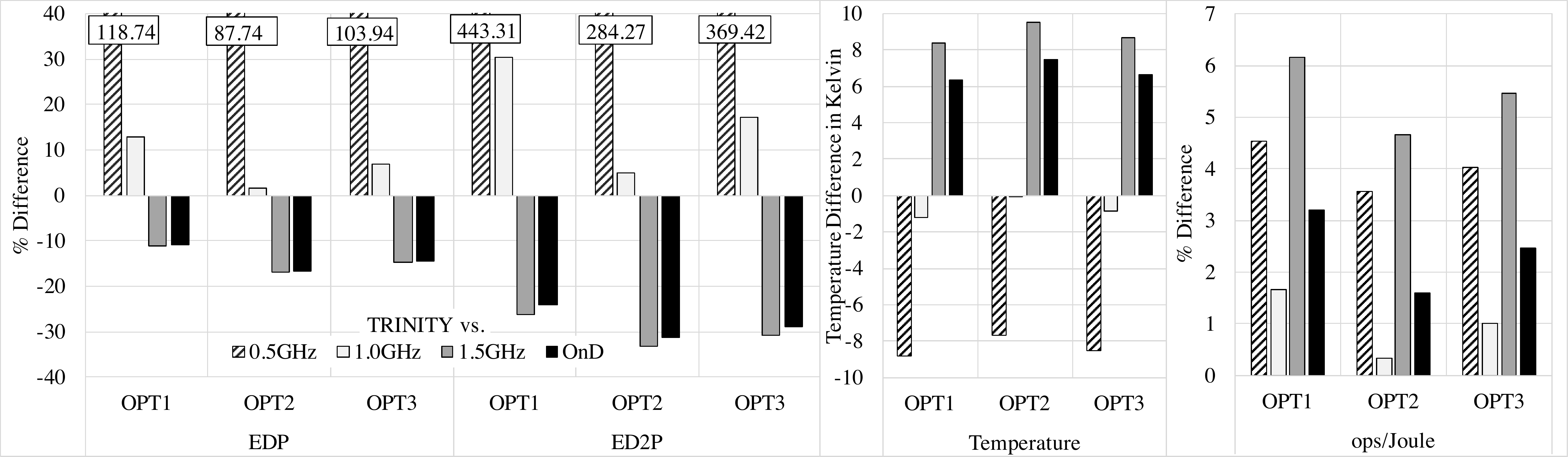}
    \vspace{-4mm}
    \caption{\textbf{EDP, ED2P, temperature and energy efficiency comparison for different TRINITY parameters}.}
    \label{fig:ParameterVariations}
    \vspace{-6mm}
\end{figure*}

On the y-axis in Figure \ref{fig:ParameterVariations} we plot the \% difference between TRINITY, \texttt{ondemand} and the three fixed frequencies over all six benchmark applications tested. We also compare temperatures of the three controller variations in the same figure. A notable feature from Fig. \ref{fig:ParameterVariations} is that with respect to $1.0$GHz, TRINITY achieves $30\%$, $5\%$ and $17\%$ better ED2P at approximately same temperature. A second feature is that as compared to $1.5$GHz and \texttt{ondemand}, TRINITY trades temperature for performance; on an average TRINITY keeps the cores at least $6$ Kelvin cooler. The large improvement in EDP and ED2P in comparison with $0.5$GHz is due to the significant reduction in the total run-time of the application. 

The third feature concurs with intuition: Increasing $\mathds{T}_R$ implies tuning $R$ less frequently thereby making the controller less responsive to changes in application phases. This increases the effect of performance mispredictions and thus reduces EDP, ED2P and ops/Joule. Consequently, the average temperature for OPT2 is higher than OPT1. Finally, decreasing $\mathds{T}$ (OPT3) does not show appreciable improvement in EDP or ED2P as compared to OPT1.

Akin to any practical thermal/power/energy management approach, TRINITY too faces the challenge of modeling precision vs. controller performance. Applications that would benefit from TRINITY are those that have a mixture of compute and memory bound phases because of the ability to adapt itself at run-time to maximally utilize the EHC. However, if those phases are shorter than the control interval $\mathds{T}$, they might end up being overlooked. TRINITY works particularly well for memory intensive applications like GraphBig because at the same EDP, the average temperature and voltage is lower than \texttt{ondemand} which improves MTTF by $68\%$. If the objective is to maximize performance alone, we get limited returns in EDP and ED2P from TRINITY. Figure \ref{fig:ParameterVariations} shows that as compared to $1.5$GHz, EDP, ED2P and ops/Joule are the lowest for TRINITY.

\subsection{Effect of Starting Temperature}
\label{subsec:StartingTemperature}

\begin{figure*}[!htb]
    \centering
    \includegraphics[scale=0.7]{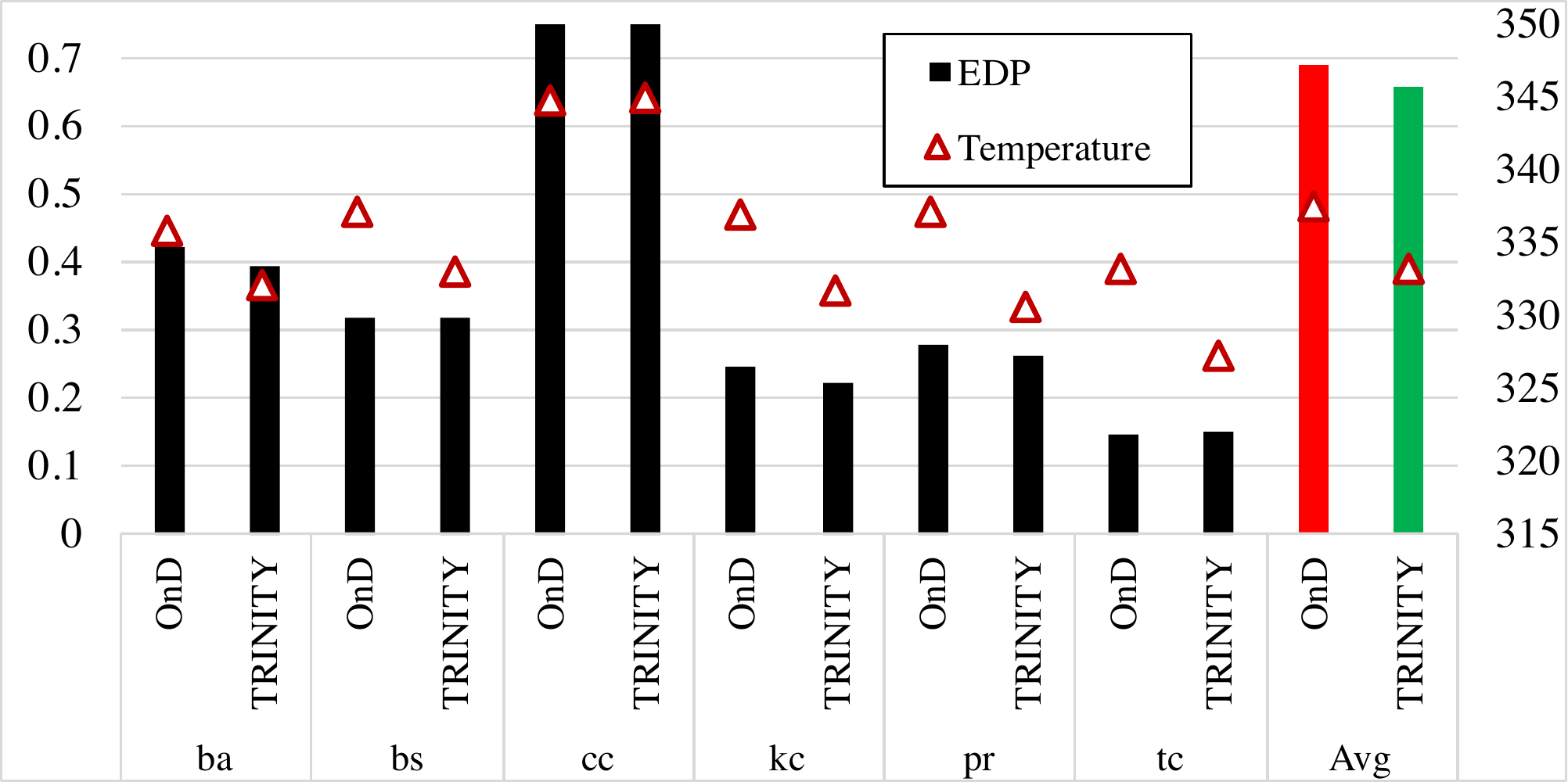}
    \caption{Controller performance compared against the \texttt{ondemand} heuristic. Controller Parameters: $T=1$ms and $T_R=1$ms. Left y-axis and right y-axis units are EDP and Kelvin, respectively.}
    \label{fig:chapter8OnDvs1msRun1EDP_Temperature}
\end{figure*}

The analysis presented in Section \ref{subsec:TRINITYPerf1} considers a starting temperature of 300K for every experiment. This however does not expose the secondary effects due to thermal throttling. To simulate practical runtime conditions, in this section we initialize the starting temperature to 323K. The corresponding EDP results are plotted in Figure \ref{fig:chapter8OnDvs1msRun1EDP_Temperature} comparing TRINITY against the \texttt{ondemand} heuristic. 

The trend of EDP is not the same for every benchmark. Consequently, the strategy to balance performance, energy and temperature should be different. For compute intensive workloads like \texttt{blackscholes} and \texttt{barnes}, the highest clock frequency (1.5GHz) delivers the best performance but also results in the highest temperature. This causes thermal throttling which significantly reduces performance. TRINITY on the other hand, tries to trade performance for benefits in temperature, 4 K on average with respect to \texttt{ondemand}. Energy efficiency results however reveal that TRINITY and \texttt{ondemand} perform equally well; TRINITY is $6.4\%$ better, arguably within simulation error bounds. The implication is indeed in line with the definition of EMT. TRINITY chooses the clock frequencies such that same or better performance can be achieved at a much lower temperature.  

Analyzing memory intensive benchmarks (\texttt{kcore}, \texttt{pagerank}, \texttt{connectedcomponent} and \texttt{tc}), there is no appreciable improvement in performance (MIPS) as core frequencies are increased. For example, average MIPS for \texttt{kcore} at $0.5$, $1.0$ and $1.5$GHz is $4360.3$, $5074.2$ and $5265.2$, respectively. Possessing \textit{apriori} knowledge that the application to be executed is memory intensive, could lead to choosing a lower clock frequency as a possible strategy. While it certainly keeps the entire 3D stack at a lower temperature, EDP can suffer considerably. Although the average power is small, the application takes much longer to complete. In these situations, TRINITY tunes $R$ in such a way that the lower half of the clock frequencies ($0.5$ - $1.0$GHz) are chosen in the memory intensive phases. Except for \texttt{connectedcomponent}, temperature of the core layer for the remaining three workloads is lower by about 6 K. For \texttt{connectedcomponent}, both \texttt{ondemand} and TRINITY perform equally well and no appreciable temperature or EDP difference is observed. 

\begin{figure*}[!htb]
    \centering
    \includegraphics[scale=0.4]{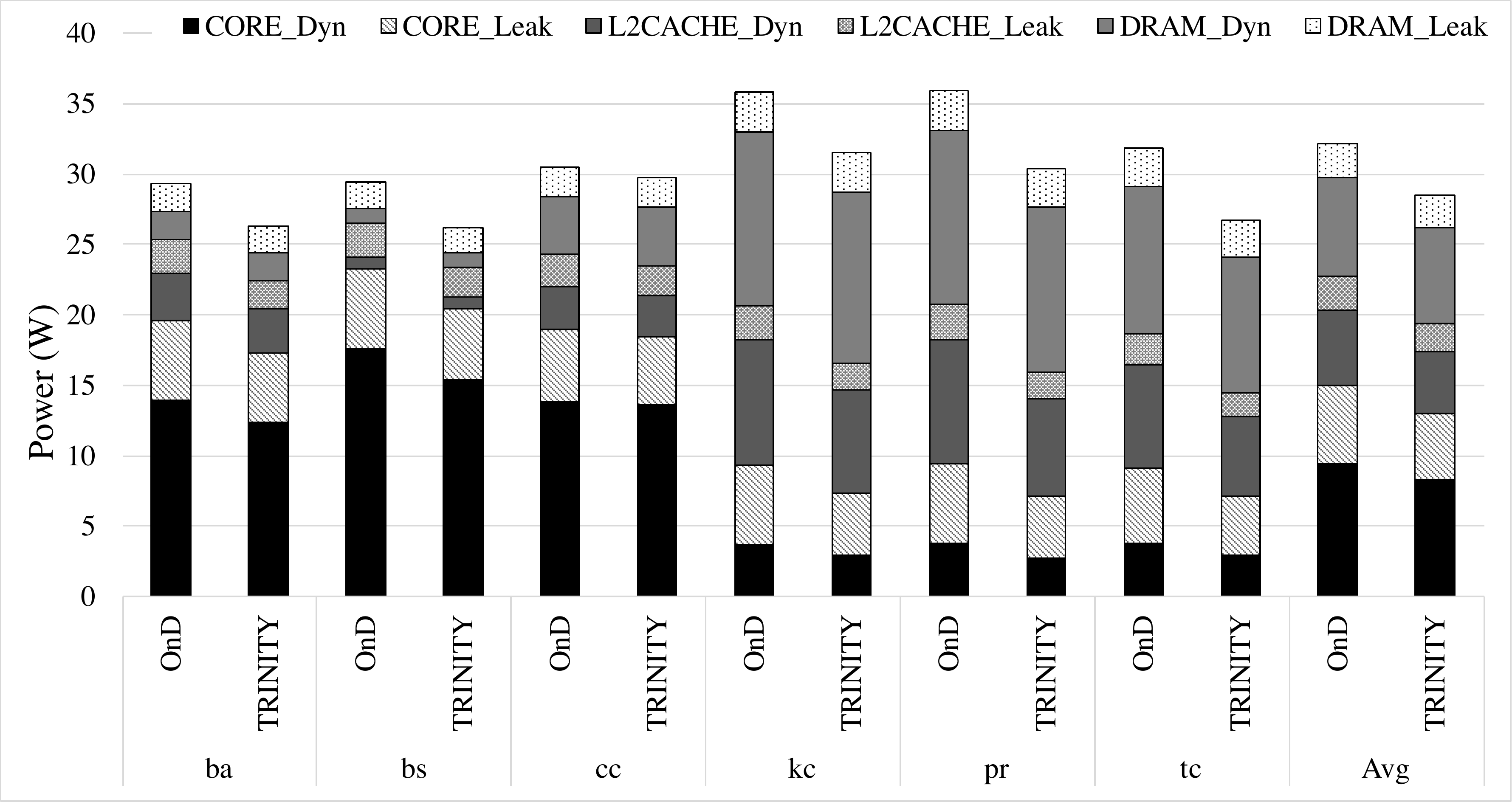}
    \caption{Average power consumption by TRINITY compared against the \texttt{ondemand} heuristic. Controller Parameters: $T=1$ms and $T_R=1$ms.}
    \label{fig:chapter8PowerComparisonOnDvsRun1_1ms}
\end{figure*}

Again, to understand the source of temperature reduction, the average power dissipated at individual layers is plotted in Figure \ref{fig:chapter8PowerComparisonOnDvsRun1_1ms}. The x-axis represents different DVFS options for each benchmark and the y-axis shows average power in Watts. Total power for each DVFS setting is broken down into dynamic and leakage power for the core, L2 Cache and DRAM layers. This distribution of power helps understand the primary source of power consumption for each benchmark application. \texttt{blackscholes} and \texttt{barnes}, both compute intensive, consume majority of the power in the core layer. \texttt{kcore}, \texttt{pagerank} and \texttt{tc} being memory intensive consume greater power in the L2 Cache layer, specifically dynamic power. Additionally, DRAM dynamic power is higher as well due to increased L2 Cache misses. \texttt{connectedcomponent}, unlike other memory bound workloads shows much higher power consumed in the core die. However, power consumed in the L2 Cache and DRAM is larger as compared to compute intensive benchmarks.  

As seen in Fig. \ref{fig:chapter8PowerComparisonOnDvsRun1_1ms}, the bulk of the power reduction (consequently reduction in temperature) comes from reducing dynamic power consumption of the core and cache layers. This is intuitive since DVFS implemented by TRINITY directly affects only the core and the corresponding L2 Cache bank. As compared to \texttt{ondemand}, dynamic power of the core and cache layers reduce by $11.7\%$ and $18\%$, respectively. Furthermore, with respect to \texttt{ondemand}, TRINITY is also able to reduce leakage power of the core and cache layers by $15.5\%$ and $16.5\%$, respectively. The power reduction can be attributed to the on-line adaptation of $R$. In memory intensive parts of the application, $\eta$ is low ($ < 1$) thus guiding the controller to choose the lower end of the clock frequencies. In compute intensive regions, $\eta$ is high ($ > 2$) allowing for higher clock frequencies to be chosen. 

TRINITY is designed so that it can be implemented on a real physical system. Simplifying the model and reducing computational complexity reduces the number of parameters that can be manually tuned. In this section, the sensitivity of TRINITY to variations in $\mathds{T}$ and $\mathds{T}_R$, which are the only manually tuned parameters, is discussed. Reducing the control cycle duration and $\mathds{T}_R$ has the benefit of capturing rapidly varying application phases. But it could also increase the amount of controller computations per unit time. Two cases are compared here: (1) OPT1 ($\mathds{T}=1$ms, $\mathds{T}_R=1$ms) and (2) OPT2 ($\mathds{T}=1$ms, $\mathds{T}_R=5$ms).

\begin{figure*}
\centering
    \begin{subfigure}[b]{0.4\textwidth}
    \includegraphics[width=\textwidth]{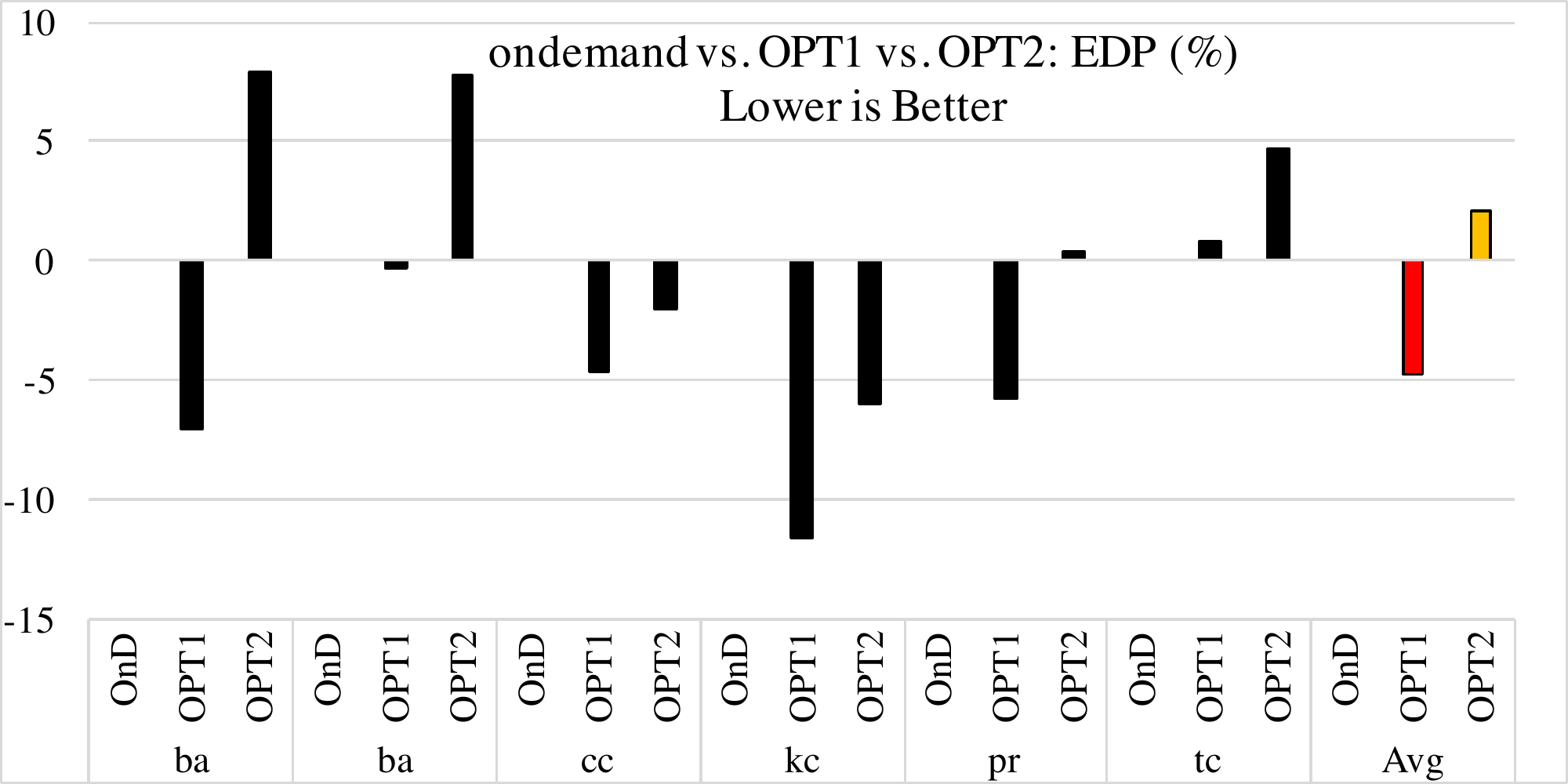}
    \caption{Comparison of EDP against \texttt{ondemand} for different TRINITY parameters.}
    \label{fig:chapter8OPT1vsOPT2_EDP}
    \end{subfigure}
    ~
    \begin{subfigure}[b]{0.4\textwidth}
    \includegraphics[width=\textwidth]{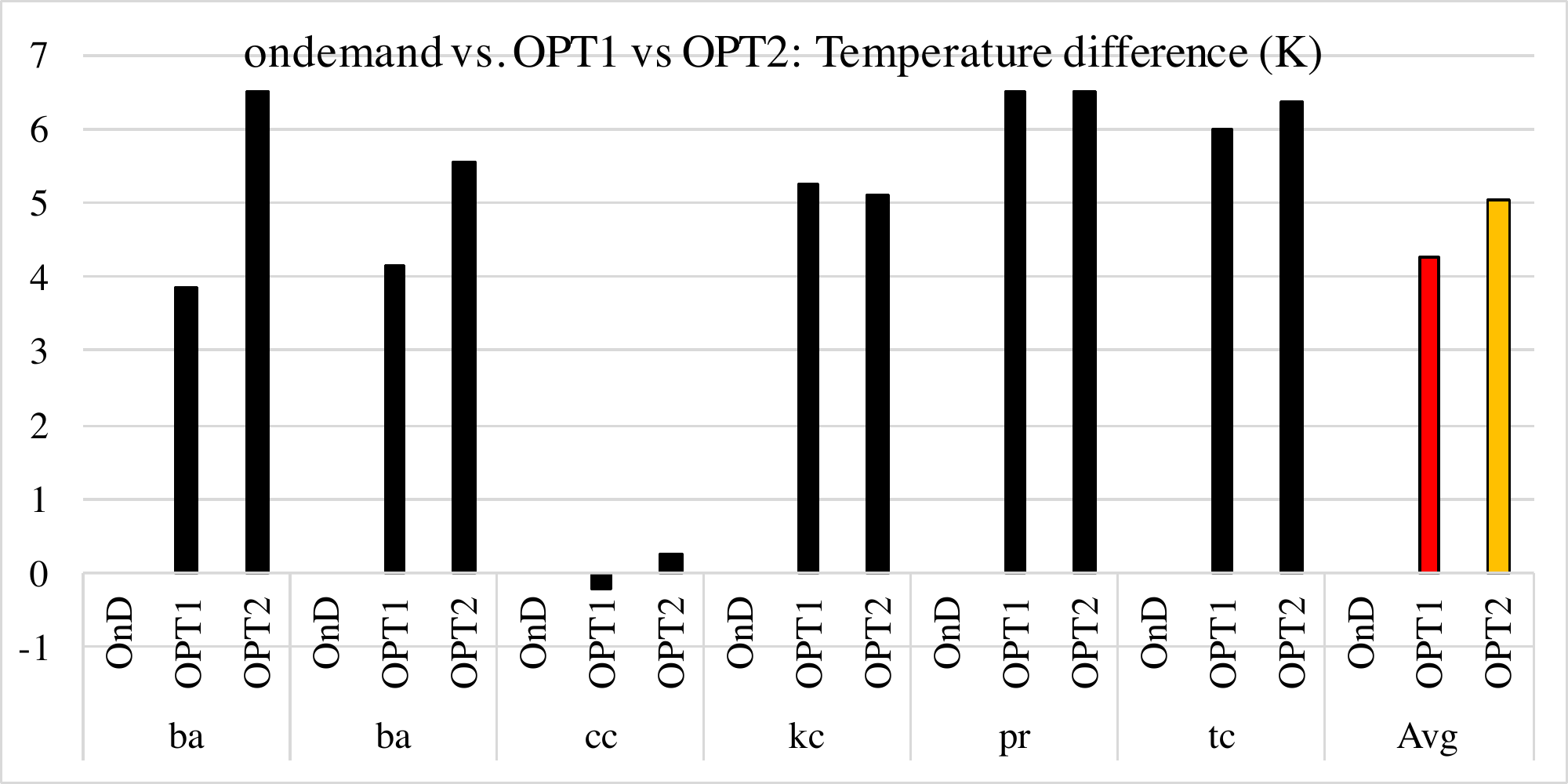}
    \caption{Comparison of Temperature against \texttt{ondemand} for different TRINITY parameters.}
    \label{fig:chapter8OPT1vsOPT2_T}
    \end{subfigure}
    
    \begin{subfigure}[b]{0.4\textwidth}
    \includegraphics[width=\textwidth]{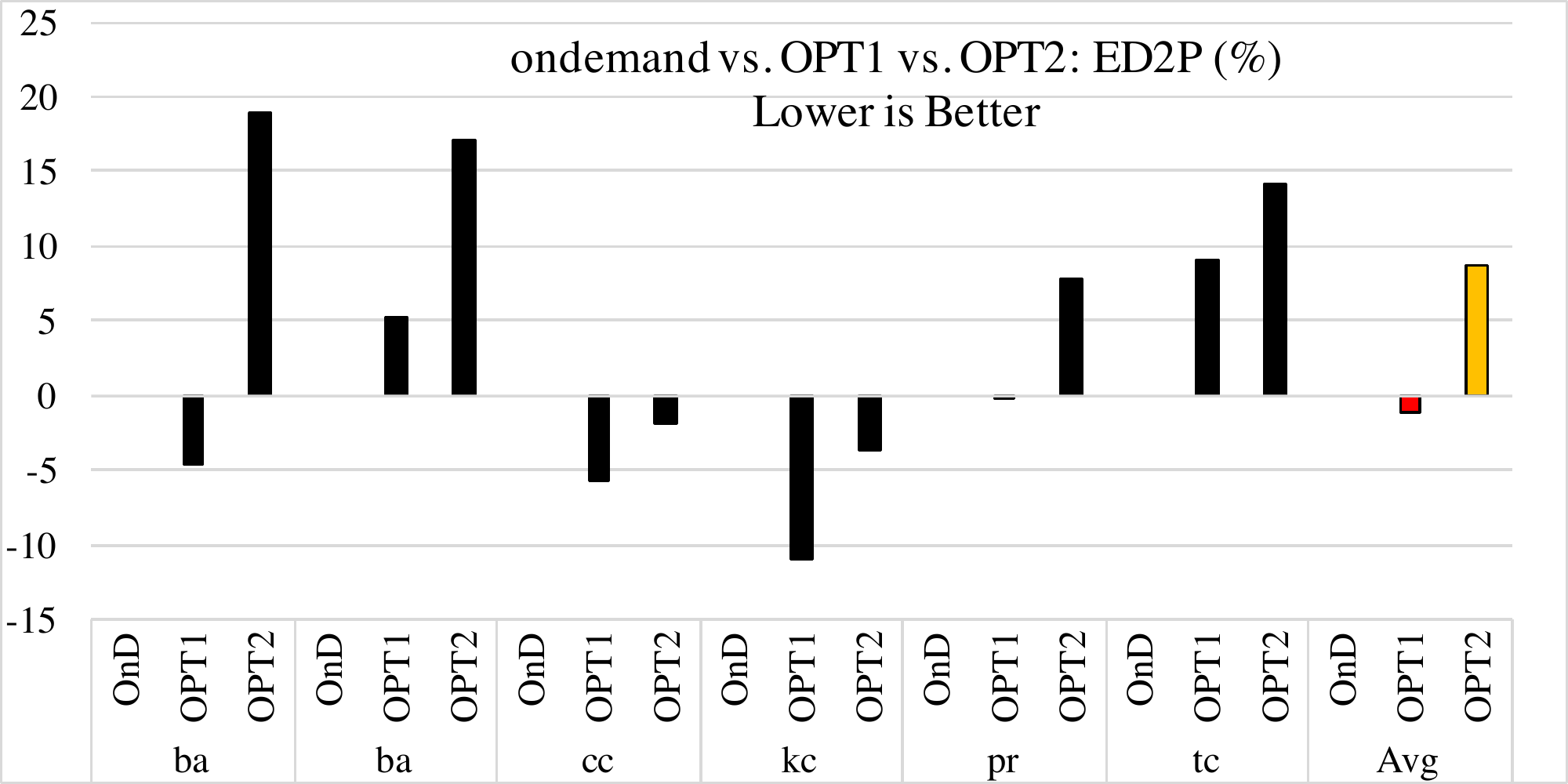}
    \caption{Comparison of ED2P against \texttt{ondemand} for different TRINITY parameters.}
    \label{fig:chapter8OPT1vsOPT2_ED2P}
    \end{subfigure}
    ~
    \begin{subfigure}[b]{0.4\textwidth}
    \includegraphics[width=\textwidth]{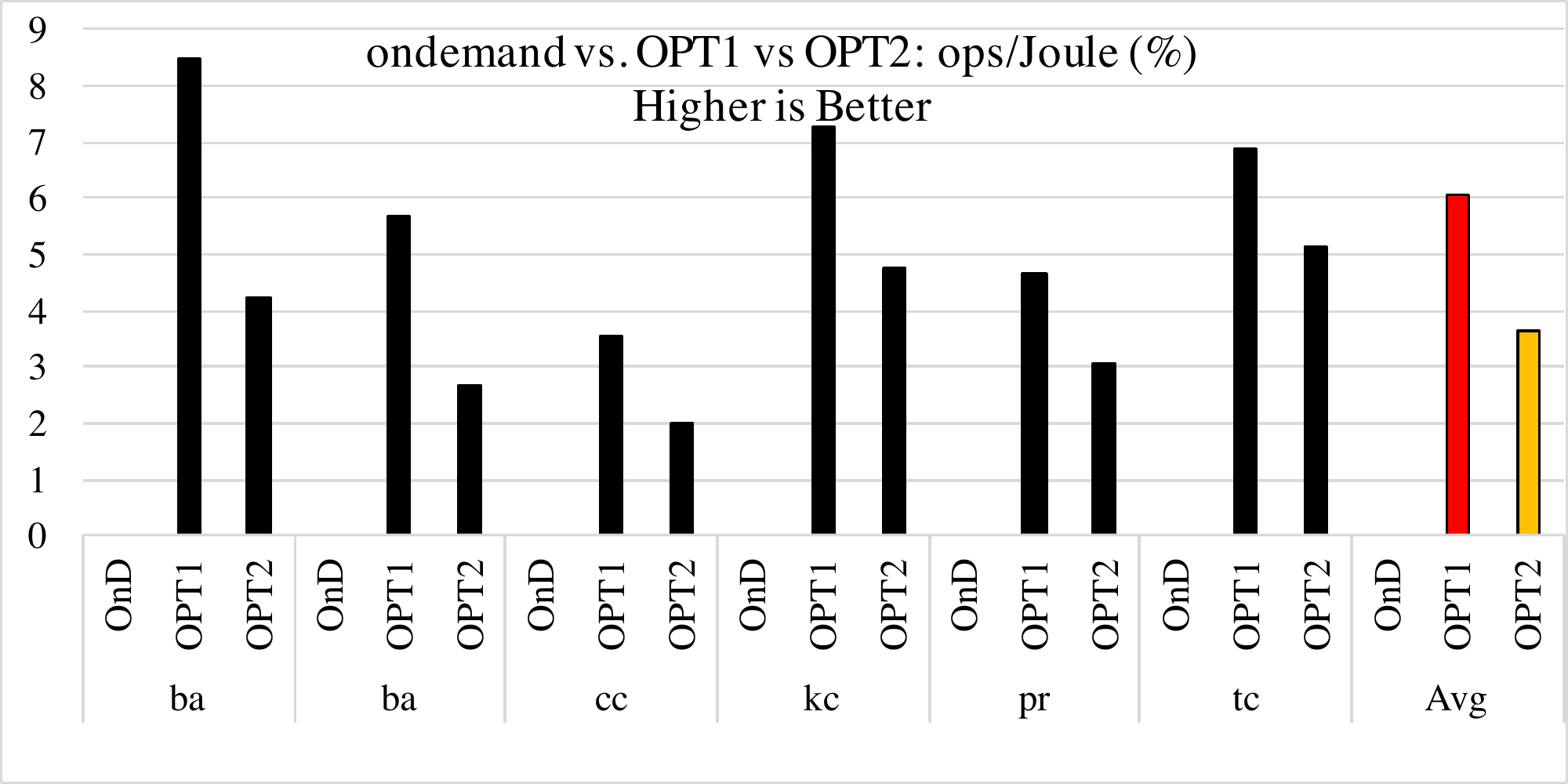}
    \caption{Comparison of ops/Joule against \texttt{ondemand} for different TRINITY parameters.}
    \label{fig:chapter8OPT1vsOPT2_opsJoule}
    \end{subfigure}
    \label{fig:chapter8ParameterVariations}
    \caption{TRINITY Parameter Variation}
\end{figure*}

The y-axis in Figure \ref{fig:chapter8OPT1vsOPT2_T} represents the absolute difference between TRINITY and \texttt{ondemand} whereas the y-axis in Figures \ref{fig:chapter8OPT1vsOPT2_EDP}, \ref{fig:chapter8OPT1vsOPT2_opsJoule} and \ref{fig:chapter8OPT1vsOPT2_ED2P} represent \% difference. Overall, OPT2 fares slightly worse than OPT1 when using the metrics EDP, ED2P and ops/Joule. OPT2 keeps the temperature of the cores about a degree cooler than OPT1 by trading off performance. This aspect is clearly observed in Figures \ref{fig:chapter8OPT1vsOPT2_EDP} and \ref{fig:chapter8OPT1vsOPT2_T}. 


The notable feature concurs with intuition: Increasing $\mathds{T}_R$ implies tuning $R$ less frequently thereby making the controller less responsive to changes in application phases. This increases the effect of performance mispredictions and thus reduces EDP, ED2P and ops/Joule. Consequently, the average temperature for OPT2 is higher than OPT1. 

Akin to any practical thermal/power/energy management approach, TRINITY too faces the challenge of modeling precision vs. controller performance. Applications that would benefit from TRINITY are those that have a mixture of compute and memory bound phases because of the ability to adapt itself at run-time to maximally utilize the EHC. However, if those phases are shorter than the control interval $\mathds{T}$, they might end up being overlooked. TRINITY works particularly well for memory intensive applications like GraphBig because at the same EDP, the average temperature and voltage is lower than \texttt{ondemand} which improves MTTF by $10\%$. 

\section{Related Work}
\label{sec:RelatedWork}
Dynamic Thermal Management (DTM) of multicore processors (2D and 3D) has evolved from heuristic based approaches to formal feedback control based techniques. They can be divided into the following categories: (i) Hardware level and (ii) Software level. At the hardware level, DVFS of independent voltage islands has been explored in great detail, for example  \cite{meng2012optimizing,zhu2008three,rao2015temperature,bartolini2013thermal,donald2006techniques,wang2009temperature,ondemand2006,zanini2010online,murali2008temperature,sahin2015just,hanumaiah2014steam}. Recently, \cite{agrawal2017xylem} discussed use of thermal TSVs to extract heat from the different layers in an architecture similar to the one described in this work. They boost the performance of applications by exploiting the improved cooling efficiency. Some other approaches to mitigate thermal effects are instruction fetch throttling, clock gating \cite{skadron2002control,skadron2003temperature} and moving the hottest datapaths closer to the heat sink (thermal herding) \cite{puttaswamy2007thermal}. Finally, at the software level we have migrating threads from hot core to cool cores \cite{yeo2008predictive,liu2012neighbor}, data compression at the memory controller \cite{khurshid2013data}, two level prefetching with throttling off-chip memory links \cite{ahn2014dynamic}, dynamic page allocation \cite{lo2016thermal}, and data block reallocation with heterogeneous memory architectures \cite{tran2013heterogeneous}. 

All the works listed in the previous paragraph except \cite{rao2015temperature} and \cite{sahin2015just} design their techniques wherein core or die temperature is an upper limit i.e. a constraint. Some policies are triggered \textit{only} under emergencies \cite{skadron2002control,skadron2003temperature}, while the rest optimize a cost while staying within the thermal threshold. TRINITY is not designed to wait till an emergency is triggered. On the contrary, it tries to avoid such scenarios by trading reduction in performance. References \cite{bartolini2013thermal,wang2009temperature,zanini2010online} develop an MPC based optimal control problem, while \cite{murali2008temperature,hanumaiah2014steam} present a numerical optimization based approach. The objectives of the problems considered vary: minimizing power \cite{wang2009temperature,murali2008temperature}, maximizing performance \cite{zhu2008three,bartolini2013thermal,zanini2010online} and maximizing power efficiency \cite{hanumaiah2014steam}. In contrast to optimization, references \cite{donald2006techniques} and \cite{rao2015temperature} design closed loop feedback controllers to maintain a fixed temperature. TRINITY is similar to the optimization based DVFS approaches but it offers an alternative approach to handle temperature and is computationally cheap to implement. Instead of considering temperature as a constraint, optimizing a cost based on it enables TRINITY to balance parameters which were otherwise dealt with in isolation. MPC and optimization based approaches referenced here are computationally intensive due to the large model dimensions and do not consider the leakage power dependence on temperature. Although \cite{bartolini2013thermal} proposes a distributed approach to MPC in an attempt to make it practically feasible, their policy eventually is heuristic based (also noted by \cite{hanumaiah2014steam}).

The work in \cite{sahin2015just} claims that minimizing thermal impact extends the sustainability of desired Quality-of-Service levels on mobile devices. In a similar vein, our work too alludes to extending the lifetime reliability of 3D stacks by keeping temperatures lower. Unlike prior works employing DVFS, we present an analysis on lifetime reliability and demonstrate the additional benefits obtained by efficiently managing the heat capacity.

\section{Conclusions}
\label{sec:Conclude}
This work presents an approach to the coordinated control of performance, energy efficiency and temperature on 3D processor-memory stacks. It introduces the concept of effective heat capacity as a thermal resource to be managed. Through a comprehensive simulation-based characterization of intra- and inter-die thermal coupling effects, the ability to maximally utilize the effective heat capacity is illustrated. An on-line DVFS controller called TRINITY is developed for the same. Unlike prior research efforts which (i) consider power, performance and temperature in isolation or in pairs, (ii) do not explicitly model static power, (iii) are heuristic based, this work acknowledges the complex interplay between performance, energy, temperature, microarchitectural parameters and package physical constraints. An analysis of EDP, ED2P, energy efficiency, temperature and also lifetime reliability is presented demonstrating the benefits of intelligently managing temperature as a resource and not just a constraint.

\section{Acknowledgements}
This work was supported in part by the National Science Foundation under Grant CNS 0855110 and the Oak Ridge National Laboratory

\bibliographystyle{ieeetr}
\bibliography{main}

\end{document}